\documentclass[a4paper,11pt]{article}

\usepackage{jheppub} 

\usepackage[T1]{fontenc} 
\usepackage{amsmath,amsthm}
\usepackage{amssymb,amsfonts}
\usepackage{graphicx}
\usepackage{dsfont}
\usepackage{relsize}
\usepackage{stmaryrd}
\usepackage{lmodern}
\usepackage{slantsc}
\usepackage{scalefnt}
\usepackage{subfigure}
\usepackage{xspace}
\usepackage{boxedminipage}
\usepackage{ifpdf}
\usepackage{multirow}
\usepackage{xifthen}
\usepackage{tensor}
\usepackage{array}
\usepackage{tabu}
\usepackage{tikz}
\usepackage{color}
\usepackage{diagbox}
\usetikzlibrary{arrows,matrix,calc,scopes,decorations.markings}
\allowdisplaybreaks[1]
\usepackage[mathcal]{euscript}
\newtheorem{theorem}{Theorem}

\newtheorem*{ansatz*}{Ansatz}
\newtheorem{principle}[theorem]{Generalized Pauli Exclusion Principle}
\newcommand{\be}{\begin{equation}}
\newcommand{\ee}{\end{equation}}
\newcommand{\bse}{\begin{subequations}}
\newcommand{\ese}{\end{subequations}}

\newcommand{\Z}{\mathbb{Z}}

\newcommand{\N}{\mathcal{N}}

\newcommand{\C}{\mathcal{C}}

\newcommand{\bpm}{\begin{pmatrix}}
\newcommand{\epm}{\end{pmatrix}}
\newcommand{\bmm}{\begin{matrix}}
\newcommand{\emm}{\end{matrix}}

\newcommand{\BLvert}{\Biggl\vert\bmm} 
\newcommand{\Brangle}{\emm\Biggr\rangle}

\newcommand{\x}{\times}
\newcommand{\ox}{\otimes}

\newcommand{\Ntbt}{N_{\tau\bar\tau}}
\tikzset{snake it/.style={decorate, decoration={snake,amplitude=0.15mm,segment length=1mm}}}
\tikzset{->-/.style={decoration={
                        markings,
                        mark=at position .55 with {\arrow{latex}}},postaction={decorate}}}
\newcommand{\vertex}{
\begin{tikzpicture}[scale=1]
\draw [->-] (5.6,9.6) -- (5.6,10.2);
\draw [->-] (5.0,9.0) -- (5.6,9.6);
\draw [->-] (6.2,9.0) -- (5.6,9.6);
\node [rotate=0.] at (6.,9.4) {\scriptsize $j$};
\node [rotate=0.] at (5.2,9.4) {\scriptsize $i$};
\node [rotate=0.] at (5.8,10.) {\scriptsize $k$};
\end{tikzpicture}
}
\newcommand{\plqt}{
\begin{tikzpicture}[scale=1]
\draw [->-] (4.8,10.4) -- (4.4,9.8);
\draw [->-] (4.4,9.8) -- (4.,10.4);
\draw [->-] (3.6,10.8) -- (4.,10.4);
\draw [->-] (5.2,10.8) -- (4.8,10.4);
\draw [->-] (4.4,9.2) -- (4.4,9.8);
\draw [->-] (4.,10.4) -- (4.8,10.4);
\node [rotate=0.] at (4.4,10.6) {\scriptsize $j_{1}$};
\node [rotate=0.] at (4.8,10.) {\scriptsize $j_{2}$};
\node [rotate=0.] at (4.,10.) {\scriptsize $j_{3}$};
\node [rotate=0.] at (5.2,10.6) {\scriptsize $j_{5}$};
\node [rotate=0.] at (4.2,9.4) {\scriptsize $j_{6}$};
\node [rotate=0.] at (3.8,10.8) {\scriptsize $j_{4}$};
\end{tikzpicture}
}
\newcommand{\plqtp}{
\begin{tikzpicture}[scale=1]
\draw [->-] (4.8,10.4) -- (4.4,9.8);
\draw [->-] (4.4,9.8) -- (4.,10.4);
\draw [->-] (3.6,10.8) -- (4.,10.4);
\draw [->-] (5.2,10.8) -- (4.8,10.4);
\draw [->-] (4.4,9.2) -- (4.4,9.8);
\draw [->-] (4.,10.4) -- (4.8,10.4);
\node [rotate=0.] at (4.4,10.6) {\scriptsize $j_{1}'$};
\node [rotate=0.] at (4.8,10.) {\scriptsize $j_{2}'$};
\node [rotate=0.] at (4.,10.) {\scriptsize $j_{3}'$};
\node [rotate=0.] at (5.2,10.6) {\scriptsize $j_{5}$};
\node [rotate=0.] at (4.2,9.4) {\scriptsize $j_{6}$};
\node [rotate=0.] at (3.8,10.8) {\scriptsize $j_{4}$};
\end{tikzpicture}
}
\newcommand{\bvertex}{
\begin{tikzpicture}[scale=1]
\path [fill, opacity=0.3] (4.2,10.8) -- (4.6,10.8) -- (4.6,9.6) -- (4.2,9.6) -- cycle;
\draw [->-] (4.6,9.6) -- (4.6,10.2);
\draw [->-] (4.6,10.2) -- (4.6,10.8);
\draw [->-] (4.6,10.2) -- (5.2,10.2);
\node [rotate=0.] at (4.4,10.6) {\scriptsize $j_{1}$};
\node [rotate=0.] at (4.4,9.8) {\scriptsize $j_{2}$};
\node [rotate=0.] at (5.,10.4) {\scriptsize $a_n$};
\end{tikzpicture}
}
\newcommand{\bplqt}{
\begin{tikzpicture}[scale=1]
\path [fill, opacity=0.3] (4.,11.2) -- (3.4,11.2) -- (3.4,8.8) -- (4.,8.8) -- (4.2,9.4) -- (4.,10.) -- (4.2,10.6) -- cycle;
\draw [->-] (4.,8.8) -- (4.2,9.4);
\draw [->-] (4.2,9.4) -- (4.,10.);
\draw [->-] (4.,10.) -- (4.2,10.6);
\draw [->-] (4.2,10.6) -- (4.,11.2);
\draw [->-] (3.6,10.) -- (4.,10.);
\draw [->-] (4.2,10.6) -- (4.8,10.6);
\draw [->-] (4.2,9.4) -- (4.8,9.4);
\node [rotate=0.] at (4.2,9.) {\scriptsize $j_{1}$};
\node [rotate=0.] at (4.2,9.8) {\scriptsize $j_{2}$};
\node [rotate=0.] at (4.,10.4) {\scriptsize $j_{3}$};
\node [rotate=0.] at (4.,11.) {\scriptsize $j_{4}$};
\node [rotate=0.] at (3.6,10.2) {\scriptsize $j_{5}$};
\node [rotate=0.] at (4.6,9.6) {\scriptsize $a_{1}$};
\node [rotate=0.] at (4.6,10.8) {\scriptsize $a_{2}$};
\end{tikzpicture}
}
\newcommand{\bplqtp}{
\begin{tikzpicture}[scale=1]
\path [fill, opacity=0.3] (4.,11.2) -- (3.4,11.2) -- (3.4,8.8) -- (4.,8.8) -- (4.2,9.4) -- (4.,10.) -- (4.2,10.6) -- cycle;
\draw [->-] (4.,8.8) -- (4.2,9.4);
\draw [->-] (4.2,9.4) -- (4.,10.);
\draw [->-] (4.,10.) -- (4.2,10.6);
\draw [->-] (4.2,10.6) -- (4.,11.2);
\draw [->-] (3.6,10.) -- (4.,10.);
\draw [->-] (4.2,10.6) -- (4.8,10.6);
\draw [->-] (4.2,9.4) -- (4.8,9.4);
\node [rotate=0.] at (4.2,9.) {\scriptsize $j_{1}$};
\node [rotate=0.] at (4.2,9.8) {\scriptsize $j'_{2}$};
\node [rotate=0.] at (4.,10.4) {\scriptsize $j'_{3}$};
\node [rotate=0.] at (4.,11.) {\scriptsize $j_{4}$};
\node [rotate=0.] at (3.6,10.2) {\scriptsize $j_{5}$};
\node [rotate=0.] at (4.6,9.6) {\scriptsize $a'_{1}$};
\node [rotate=0.] at (4.6,10.8) {\scriptsize $a'_{2}$};
\end{tikzpicture}
}

\newcommand{\boxttbar}{
\begin{tikzpicture}[scale=1,baseline=303.4]
\draw [dashed, thick] (0.8,11.2) -- (1.2,11.2);
\draw [thick] (0.8,11.2) -- (0.8,10.8);
\draw [thick] (0.8,10.8) -- (1.2,10.8);
\draw [thick] (0.8,11.2) -- (1.2,11.2);
\draw [thick] (1.2,11.2) -- (1.2,10.8);
\draw [dashed, thick] (0.8,10.7) -- (1.2,10.7);
\draw [thick] (0.8,10.7) -- (0.8,10.3);
\draw [thick] (0.8,10.3) -- (1.2,10.3);
\draw [thick] (0.8,10.7) -- (1.2,10.7);
\draw [thick] (1.2,10.7) -- (1.2,10.3);
\end{tikzpicture}
}

\newcommand{\boxone}{
\begin{tikzpicture}[scale=1,baseline=303.4]
\draw [dashed, thick] (0.8,11.2) -- (1.2,11.2);
\draw [thick] (0.8,11.2) -- (0.8,10.8);
\draw [thick] (0.8,10.8) -- (1.2,10.8);
\draw [thick] (0.8,11.2) -- (1.2,11.2);
\draw [thick] (1.2,11.2) -- (1.2,10.8);
\draw [dashed, thick] (0.8,10.7) -- (1.2,10.7);
\draw [thick] (0.8,10.7) -- (0.8,10.3);
\draw [thick] (0.8,10.3) -- (1.2,10.3);
\draw [thick] (0.8,10.7) -- (1.2,10.7);
\draw [thick] (1.2,10.7) -- (1.2,10.3);
\node [rotate=0.] at (1.,11.) {\scriptsize $\blacksquare$};
\node [rotate=0.] at (1.,10.5) {\scriptsize $\blacksquare$};
\end{tikzpicture}
}

\newcommand{\boxonetbar}{
\begin{tikzpicture}[scale=1,baseline=303.4]
\draw [dashed, thick] (0.8,11.2) -- (1.2,11.2);
\draw [thick] (0.8,11.2) -- (0.8,10.8);
\draw [thick] (0.8,10.8) -- (1.2,10.8);
\draw [thick] (0.8,11.2) -- (1.2,11.2);
\draw [thick] (1.2,11.2) -- (1.2,10.8);
\draw [dashed, thick] (0.8,10.7) -- (1.2,10.7);
\draw [thick] (0.8,10.7) -- (0.8,10.3);
\draw [thick] (0.8,10.3) -- (1.2,10.3);
\draw [thick] (0.8,10.7) -- (1.2,10.7);
\draw [thick] (1.2,10.7) -- (1.2,10.3);
\node [rotate=0.] at (1.,11.) {\scriptsize $\blacksquare$};
\end{tikzpicture}
}

\newcommand{\boxtone}{
\begin{tikzpicture}[scale=1,baseline=303.4]
\draw [dashed, thick] (0.8,11.2) -- (1.2,11.2);
\draw [thick] (0.8,11.2) -- (0.8,10.8);
\draw [thick] (0.8,10.8) -- (1.2,10.8);
\draw [thick] (0.8,11.2) -- (1.2,11.2);
\draw [thick] (1.2,11.2) -- (1.2,10.8);
\draw [dashed, thick] (0.8,10.7) -- (1.2,10.7);
\draw [thick] (0.8,10.7) -- (0.8,10.3);
\draw [thick] (0.8,10.3) -- (1.2,10.3);
\draw [thick] (0.8,10.7) -- (1.2,10.7);
\draw [thick] (1.2,10.7) -- (1.2,10.3);
\node [rotate=0.] at (1.,10.5) {\scriptsize $\blacksquare$};
\end{tikzpicture}
}

\newcommand{\twoSpeciesRule}{
\begin{tikzpicture}[baseline={([yshift=-.8ex]current bounding box.center)},scale=1]
\draw [thick] (3.4,10.8) -- (3.4,10.4);
\draw [thick] (3.4,10.4) -- (3.8,10.4);
\draw [thick] (3.8,10.4) -- (3.8,10.8);
\draw [thick] (3.4,10.8) -- (3.8,10.8);
\draw [thick] (3.4,10.9) -- (3.8,10.9);
\draw [thick] (3.8,10.9) -- (3.8,11.3);
\draw [thick] (3.8,11.3) -- (3.4,11.3);
\draw [thick] (3.4,11.3) -- (3.4,10.9);
\draw [] (3.5,10.7) -- (3.5,10.5);
\draw [] (3.5,10.5) -- (3.7,10.5);
\draw [] (3.7,10.5) -- (3.7,10.7);
\draw [] (3.7,10.7) -- (3.5,10.7);
\draw [thick] (2.4,10.8) -- (2.4,10.4);
\draw [thick] (2.4,10.4) -- (2.8,10.4);
\draw [thick] (2.8,10.4) -- (2.8,10.8);
\draw [thick] (2.4,10.8) -- (2.8,10.8);
\draw [thick, dotted] (2.4,11.3) -- (2.4,10.9);
\draw [thick, dotted] (2.4,10.9) -- (2.8,10.9);
\draw [thick, dotted] (2.8,10.9) -- (2.8,11.3);
\draw [thick, dotted] (2.4,11.3) -- (2.8,11.3);
\draw [] (2.5,10.7) -- (2.5,10.5);
\draw [] (2.5,10.5) -- (2.7,10.5);
\draw [] (2.7,10.5) -- (2.7,10.7);
\draw [] (2.5,10.7) -- (2.7,10.7);
\node [rotate=0.] at (3.1,10.8) {\scriptsize $=$};
\node [rotate=0.] at (3.6,11.1) {\scriptsize $\x$};
\end{tikzpicture}
}

\newcommand{\AnyonAndBdries}{
\begin{tikzpicture}[baseline={([yshift=-.8ex]current bounding box.center)},scale=0.6]
\draw [] (5.2,9.1) ..controls (5.273,9.245) and (5.527,9.391) .. (5.8,9.3)
..controls (6.073,9.209) and (6.364,8.882) .. (6.4,8.7)
..controls (6.436,8.518) and (6.218,8.482) .. (6.,8.5)
..controls (5.782,8.518) and (5.564,8.591) .. (5.4,8.7)
..controls (5.236,8.809) and (5.127,8.955) .. (5.2,9.1);
\draw [] (7.2,8.6) ..controls (7.2,8.75) and (7.55,8.9) .. (7.8,8.8)
..controls (8.05,8.7) and (8.2,8.35) .. (8.2,8.2)
..controls (8.2,8.05) and (8.05,8.1) .. (7.8,8.2)
..controls (7.55,8.3) and (7.2,8.45) .. (7.2,8.6);
\draw [] (9.,9.4) ..controls (9.05,9.2) and (9.3,9.15) .. (9.4,9.2)
..controls (9.5,9.25) and (9.45,9.4) .. (9.4,9.6)
..controls (9.35,9.8) and (9.3,10.05) .. (9.2,10.)
..controls (9.1,9.95) and (8.95,9.6) .. (9.,9.4);
\draw [] (10.,10.4) ..controls (9.8,10.35) and (9.45,10.45) .. (9.4,10.6)
..controls (9.35,10.75) and (9.6,10.95) .. (9.8,11.)
..controls (10.,11.05) and (10.15,10.95) .. (10.2,10.8)
..controls (10.25,10.65) and (10.2,10.45) .. (10.,10.4);
\draw [lightgray, fill] (5.2,9.1) ..controls (5.273,9.245) and (5.527,9.391) .. (5.8,9.3)
..controls (6.073,9.209) and (6.364,8.882) .. (6.4,8.7)
..controls (6.436,8.518) and (6.218,8.482) .. (6.,8.5)
..controls (5.782,8.518) and (5.564,8.591) .. (5.4,8.7)
..controls (5.236,8.809) and (5.127,8.955) .. (5.2,9.1);
\draw [lightgray, fill] (7.2,8.6) ..controls (7.2,8.75) and (7.55,8.9) .. (7.8,8.8)
..controls (8.05,8.7) and (8.2,8.35) .. (8.2,8.2)
..controls (8.2,8.05) and (8.05,8.1) .. (7.8,8.2)
..controls (7.55,8.3) and (7.2,8.45) .. (7.2,8.6);
\draw [lightgray, fill] (9.,9.4) ..controls (9.05,9.2) and (9.3,9.15) .. (9.4,9.2)
..controls (9.5,9.25) and (9.45,9.4) .. (9.4,9.6)
..controls (9.35,9.8) and (9.3,10.05) .. (9.2,10.)
..controls (9.1,9.95) and (8.95,9.6) .. (9.,9.4);
\draw [lightgray, fill] (10.,10.4) ..controls (9.8,10.35) and (9.45,10.45) .. (9.4,10.6)
..controls (9.35,10.75) and (9.6,10.95) .. (9.8,11.)
..controls (10.,11.05) and (10.15,10.95) .. (10.2,10.8)
..controls (10.25,10.65) and (10.2,10.45) .. (10.,10.4);
\draw [fill,rotate around={0.:(3.6,10.2)}] (3.6,10.2) ++(0.:0.08 and 0.08) arc(0.:360.:0.08 and 0.08);
\draw [fill,rotate around={0.:(4.3,10.5)}] (4.3,10.5) ++(0.:0.08 and 0.08) arc(0.:360.:0.08 and 0.08);
\draw [fill,rotate around={0.:(5.2,10.2)}] (5.2,10.2) ++(0.:0.08 and 0.08) arc(0.:360.:0.08 and 0.08);
\draw [fill,rotate around={0.:(2.2,9.5)}] (2.2,9.5) ++(0.:0.08 and 0.08) arc(0.:360.:0.08 and 0.08);
\draw [fill,rotate around={0.:(3.,9.5)}] (3.,9.5) ++(0.:0.08 and 0.08) arc(0.:360.:0.08 and 0.08);
\end{tikzpicture}}

\newcommand{\AnyonAndBdriesAA}{\begin{tikzpicture}[baseline={([yshift=-.8ex]current bounding box.center)},scale=0.6]
        \draw [fill,rotate around={0.:(3.6,10.2)}] (3.6,10.2) ++(0.:0.08 and 0.08) arc(0.:360.:0.08 and 0.08);
        \draw [fill,rotate around={0.:(4.3,10.5)}] (4.3,10.5) ++(0.:0.08 and 0.08) arc(0.:360.:0.08 and 0.08);
        \draw [fill,rotate around={0.:(5.2,10.2)}] (5.2,10.2) ++(0.:0.08 and 0.08) arc(0.:360.:0.08 and 0.08);
        \draw [fill,rotate around={0.:(2.2,9.5)}] (2.2,9.5) ++(0.:0.08 and 0.08) arc(0.:360.:0.08 and 0.08);
        \draw [fill,rotate around={0.:(3.,9.5)}] (3.,9.5) ++(0.:0.08 and 0.08) arc(0.:360.:0.08 and 0.08);
        \draw [gray,fill,rotate around={0.:(5.8,9.2)}] (5.8,9.2) ++(0.:0.2 and 0.2) arc(0.:360.:0.2 and 0.2);
        \draw [gray,fill,rotate around={0.:(7.8,8.7)}] (7.8,8.7) ++(0.:0.2 and 0.2) arc(0.:360.:0.2 and 0.2);
        \draw [gray,fill,rotate around={0.:(9.2,9.9)}] (9.2,9.9) ++(0.:0.2 and 0.2) arc(0.:360.:0.2 and 0.2);
        \draw [gray,fill,rotate around={0.:(9.8,10.9)}] (9.8,10.9) ++(0.:0.2 and 0.2) arc(0.:360.:0.2 and 0.2);
        \end{tikzpicture}}

\newtabulinestyle{dash=off 2pt}

 {\null\,\vcenter\bgroup\scriptsize
  \arraycolsep=.13885em
  \hbox\bgroup$\array{@{}#1@{}}}
 {\endarray$\egroup\egroup\,\null}

\newcommand{\atildedisk}{\tilde\alpha^{D^1}_{\tau\bar\tau}}
\newcommand{\atildemulti}{\tilde\alpha^{N_1}_{\tau\bar\tau}}
\newcolumntype{C}{>{\centering\arraybackslash} m{1.5em} }

\makeatletter
\newcommand*{\Relbarfill@}{\arrowfill@\Relbar\Relbar\Relbar}
\newcommand*{\xeq}[2][]{\ext@arrow 0055\Relbarfill@{#1}{#2}}
\makeatother

\title{Anyonic exclusions statistics on surfaces with gapped boundaries}
\date{\today}
\author[a,b]{Yingcheng Li}
\author[a,b]{Hongyu Wang}
\author[d]{Yuting Hu}
\author[a,b,c,d,e,1]{Yidun Wan,\note{Corresponding author}}
\affiliation[a]{State Key Laboratory of Surface Physics, Fudan University, Shanghai 200433, China}
\affiliation[b]{Department of Physics and Center for Field Theory and Particle Physics, Fudan University, Shanghai 200433, China}
\affiliation[c]{Institute for Nanoelectronic devices and Quantum computing, Fudan University, Shanghai 200433, China}
\affiliation[d]{Department of Physics and Institute for Quantum Science and Engineering, Southern University of Science and Technology, Shenzhen 518055, China}
\affiliation[e]{Collaborative Innovation Center of Advanced Microstructures, Nanjing, 210093, China}
\emailAdd{ydwan@fudan.edu.cn}

\abstract{
An anyonic exclusion statistics, which generalizes the Bose-Einstein and Fermi-Dirac statistics of bosons and fermions, was proposed by Haldane\cite{Haldane1991a}. When fusion of anyons is involved, certain `pseudo-species' anyons appear in the exotic statistical weights of non-Abelian anyon systems; however, the meaning and significance of pseudo-species remains an open problem. The relevant past studies had considered only anyon systems without any physical boundary but boundaries often appear in real-life materials. In this paper, we propose an extended anyonic exclusion statistics on surfaces with gapped boundaries, introducing mutual exclusion statistics between anyons as well as the boundary components. Motivated by Refs. \cite{Wu1994,Hu2013}, we present a formula for the statistical weight of many-anyon states obeying the proposed statistics. Taking the (doubled) Fibonacci topological order as an example, we develop a systematic basis construction for non-Abelian anyons on any Riemann surfaces with gapped boundaries. The basis construction offers a standard way to read off a canonical set of statistics parameters and hence write down the extended statistical weight of the anyon system being studied. The basis construction reveals the meaning of pseudo-species. A pseudo-species has different `excitation' modes, each corresponding to an anyon species. The `excitation' modes of pseudo-species corresponds to good quantum numbers of subsystems of a non-Abelian anyon system. This is important because often (e.g., in topological quantum computing) we may be concerned about only the entanglement between such subsystems.
}

\begin{document}

\maketitle
\flushbottom
\section{Introduction}\label{sec:intro}
Anyons are one of the most interesting properties of topologically ordered matter phases. Due to the exotic (as compared with the usual bosonic and fermionic) exchange statistics of non-Abelian anyons, certain kinds of non-Abelian
anyons can be braided to simulate quantum-gate processing, leading to a scheme of robust quantum computing---topological quantum computing\cite{Kitaev2003a,Freedman2003,Stern2006,Nayak2008}. A candidate of such anyon system is the $\nu=12/5$ fractional quantum Hall liquid\cite{Nayak2008}, which is thought to bear Fibonacci anyons. Anyons are nonlocal objects. In particular, a non-Abelian anyon does not even have a well defined local Hilbert space but carry a quantum dimension---usually an irrational number---which is the asymptotic dimension of the Hilbert space associated with an anyon in the thermodynamic limit. Hence, a qubit (or more generally a qunit) in topological quantum computing is identified as a subspace of the Hilbert space of multiple non-Abelian anyons. Such a Hilbert space is defined as the fusion space of non-Abelian anyon states and grows with the number of anyons. If we know the analytic relation between the dimension of the Hilbert space and the number of anyons, in the thermodynamic limit, we may obtain an exotic statistics of the anyons that generalizes the Bose-Einstein and Fermi-Dirac statistics. 

Two important questions then follow. 1) If such an exotic exclusion statistics exists, would it lead to exotic exclusion principles of the relevant anyons that generalizes the Pauli exclusion principle of fermions? 2) In a system of large number of non-Abelian anyons, how would we characterize the subsystems of certain anyons?

The first question has been partially answered. In Refs. \cite{Haldane1991a}, Haldane proposed a generalized Pauli principle, stating that the number of available single particle states for additional particles depends on the number of (each type of) the existing particles. Such a relation can be formulated by a linear difference equation, with the linear coefficients being defined as the exclusion statistics parameters. 
Later, in Ref. \cite{Wu1994}, Wu developed a statistical weight formula of ideal gases of anyons---an integral form of Haldane's difference equation, as an explicit generalization of that of bosons and fermions, and studied the thermodynamics of anyons. The Haldane-Wu statistics for Abelian anyons is also discussed in $1$-dimensional systems\cite{batchelor_generalized_2006,batchelor_onedimensional_2006}. When non-Abelian anyons are involved, however, to still apply the ideal gas picture and derive the thermodynamics of anyons, the original state counting of many-anyon systems by Haldane and Wu would have to be appropriately modified. Then in Ref. \cite{Hu2013}, the authors studied the statistical weights of certain non-Abelian anyons by taking into account the fusion between the anyons; they obtained a more generalized formula of statistical weights, in which the notion of pseudo-species\footnote{This notion was first introduced in Ref. \cite{Guruswamy1999a} in the context of conformal field theories.} was adopted to account for the exotic exclusion statistics due to fusion; however, unfortunately, the physical meaning of pseudo-species remains an open problem. 

Both Haldane\cite{Haldane1991a} and Wu\cite{Wu1994} considered ideal gases with a fixed boundary condition, which is in fact the periodic boundary condition, which renders the anyon systems being on spaces without actual boundaries. Neither did Ref. \cite{Hu2013} consider boundaries.

Nevertheless, materials with boundaries are much easier to fabricate than closed ones. Understanding the anyonic exclusion statistics in topologically ordered states with boundaries is thus important. For such a system to have a well-defined, topologically protected, degenerate ground-state Hilbert space, which may support a robust quantum memory and quantum computing\cite{Kitaev2003a,Kitaev2006}, the systems with gapped boundaries are of most interest. A recent work\cite{HungWan2014} has shown how gapped boundary conditions of a topological order dictate the ground state degeneracy of the topological order and how certain anyons in the bulk may connect to the gapped boundary. The fusion space structure of multiple anyons is closely related to the boundary conditions, which select only certain anyons that can move to the gapped boundary without any energy cost\footnote{In other words, these anyons condense at the boundary\cite{Kitaev2012,Levin2013,HungWan2013a,HungWan2014}.}. Hence we expect that the boundary conditions of a topological order would affect the state counting of the anyons. 

The second question regards how a subsystem of anyons in a many-anyon system may be characterized by certain good quantum numbers. This is an  important question because often (e.g., in topological quantum computing) we may be concerned about only the entanglement between such subsystems and ignore what is inside each subsystem.  Recall that the Hilbert space of a fermionic or bosonic system is taken as a Fock space, which is the tensor product of the local Hilbert spaces of single-particle states. To study the Hilbert space structure of a system of non-Abelian anyons, however, there demands a new formulation of the basis, as many-anyon states do not have an obvious Fock space analogy formed by the tensor product of local single-anyon Hilbert spaces. If we can find the complete set of observables (giving rise to good quantum numbers) of the system, we may take their eigenvectors to form the basis of the Hilbert space. As to be seen in this paper, the clarification of the physical meaning of pseudo-species turns out to be the answer to this question too. More precisely, the good quantum numbers of a subsystem are the eigenvalues of the observables of the relevant pseudo-species.

In this paper, we extend Haldane's generalized exclusion principle to systems with gapped boundaries, by proposing that the \textit{number of available single particle states for additional particles linearly and mutually depends on the number of (every species of) existing anyons and the number of (every boundary type of) existing gapped boundaries}. The dependency coefficients are called the statistics parameters. In terms of these extended statistics parameters, we extend Wu's formula for statistical weight, which lays the foundation of the statistical mechanics of anyons on systems with gapped boundaries. Clearly, such exclusion statistics put the boundary components and bulk anyons on an equal footing. We shall dub this kind of exclusion statistics by \textit{extended anyonic exclusion statistics}.

To verify the extended exclusion statistics, we compute the statistical weight in an extension\cite{Hu2017,Hu2017a}  of the Levin-Wen model\cite{Levin2004}---a discrete topological quantum field theory (TQFT) for topological phases, and derive the corresponding statistics parameters. Namely, we adopt the extended Levin-Wen model of topological orders with gapped boundaries recently constructed by two of us\cite{Hu2017,Hu2017a} , and compute the state counting of multiple anyon excitations in this model. The state counting so obtained leads to the extended exclusion statistics of the anyons under consideration. 

The computing results motivate us to develop a trustworthy universal method of basis construction of any multi-anyon Hilbert space on any Riemann surface with boundaries. Using this method of basis, we can easily read off the statistical weight being studied. The method of basis also reveals the true meaning of pseudo-species and how boundary conditions affect the pseudo-species. Our main results are as follows.
\begin{enumerate}
        \item We propose an extended anyonic exclusion statistics on an open surface with gapped boundaries, and present a formula for the statistical weight of many-anyon states obeying the proposed statistics. Our formula include pseudo-species contribution and thus works for non-Abelian anyons.
        %
        \item We develop a systematic basis construction for systems of non-Abelian anyons on a disk with a gapped boundary. From the basis construction, we can directly read off the corresponding statistics parameters and hence derive the statistical weight.
        \item The basis construction reveals the meaning of pseudo-species. A pseudo-species has different `excitation' modes, each corresponding to an anyon species.  The `excitation' modes of pseudo-species corresponds to good quantum numbers of subsystems of an anyon system.
        %
        %
        \item We introduce topological operations on the many-anyon Hilbert space. We present how to construct bases for many-anyon states on any Riemann surfaces with multiple boundaries, by applying the topological operations on the bases on a disk. This provides a standard way to read off the statistics parameters.
        
\end{enumerate}    

To convey the ideas and deliver the results more lucidly, we shall take the doubled Fibonacci topological order as a concrete example, by numerical as well as analytical computations. We make this choice also because Fibonacci anyons are the most promising and simplest anyons that can in principle realize universal topological quantum computing. We shall call a quantum system that admits the (doubled) Fibonacci topological order simply a Fibonacci system. 

Our methods---in particular our systematic basis construction may apply to all anyons systems describable by Reshetikhin-Turaev topological field theories. Doing so will reveal deeper significance of anyonic exclusion statistics. The results are to be reported in a companion paper\footnote{In preparation.}.

The paper is structured as follows. Section \ref{sec:rev} introduces our extended anyonic exclusion statistics, accompanied by briefing of the concept of anyon exclusions statistics. Section \ref{sec:Disk} computes the state counting and derive the statistical weight of doubled Fibonacci anyons on a disk with a gapped boundary, using the extended Levin-Wen model. Motivated by the results of Section \ref{sec:Disk}, Section \ref{sec:basisDisk} systematically constructs the bases of the Hilbert spaces of Fibonacci system on a disk, from which the statistics parameters can be immediately read off. Section \ref{sec:pseudo} addresses the physical meaning of pseudo-species and the boundary effect on state counting and pseudo-species. Section \ref{sec:multi} generalizes the story to surfaces with multiple gapped boundaries by two topological operations. Section \ref{sec:equiv} brings up an important equivalence relation between statistical weight and fusion basis. Appendices collect certain reviews, details, and more examples, including an Abelian example---the $\Z_2$ toric code order, which further corroborate our results. 

\section{Extended anyonic exclusion statistics}\label{sec:rev}
In two spatial dimensions, anyons have exotic exchange statistics, namely, exchanging two anyons brings a phase factor---other than $\pm\ 1$---to the system's wavefunction. Anyons also have exotic exclusion statistics, which was first studied even before the notion of topological orders was proposed, leading to a generalized Pauli exclusion principle\cite{Haldane1991a,Wu1994,Hu2013}. Unlike the exotic exchange statistics of anyons\footnote{Here we refer to pointlike quasiparticles only. Higher dimensional quasiparticles in higher dimensional topological orders may still possess exotic exchange statistics (see Ref.\cite{Wan2014} for an example).}, which makes sense only in two spatial dimensions, the exotic exclusion statistics makes sense even in one spatial dimension, where exchanging two anyons becomes unphysical\footnote{One may claim that exchanging two anyons in $1$D is possible by considering exchaning two anyons on the $1$D boundary of a $2$D bulk theory via the bulk; however, this is beyond our concern.}. Anyonic exclusion statistics was first proposed by Haldane\cite{Haldane1991a} by writing down a linear difference equation for state counting.  Later, Wu explicitly wrote down an integral form of the generalized statistical weight \eqref{eq:StatWeightMulti} of anyons. This generalized statistical weight encompasses the usual bosons and fermions, and more importantly, works not only in the thermodynamical limit but also in the case of finite number of anyons. 

\subsection{Review of exotic exclusion statistics of anyons}
The exclusion statistics of $N$ anyons is determined by the number of states permitted by the system for these anyons to coexist. Consider an ideal gas of $N$ anyons of the same species and period boundary condition applied. According to Refs.\cite{Haldane1991a,Wu1994}, the exclusion statistics of this ideal gas is captured by the statistical weight $W_{G,N}$: 
\be\label{eq:StatWeight1}
W_{G,N}=\bpm
G_{eff}+(N-1)\\
N
\epm,
\ee  
where $G$ is the number of single-particle states for $N=1$, and $G_{eff}=G-\alpha(N-1)$ being the number of available single-particle states. Here, $\alpha$ is called the statistics parameter between the anyons. In the case of bosons, $\alpha=0$, and $\alpha=1$ for fermions. If the ideal gas contains multiple species of anyons, say, $\{G_i,N_i\}$, where $N_i$ is the anyon number of the $i$-th species, and $G_i$ the number of single particle state when $N_i=1$. Then, the generalized statistical weight reads 
\be\label{eq:StatWeightMulti}
W_{\{G_i,N_i\}}=\prod_{i=0}^{n-1}\bpm
G_i+N_i-1-\sum_{j=0}^{n-1}\alpha_{ij}(N_j-\delta_{ij})\\
\\
N_i
\epm.
\ee
Here $n$ denotes the total number of anyon species. The generalization lies in that the statistics parameter $\alpha$ is now a matrix, whose matrix element $\alpha_{ij}$ is the statistics parameter between the $i$-th and $j$-th species of anyons. 

Since in an ideal gas, one ignores the particle-like interaction---that is the fusion---of anyons, the matrix $\alpha$ in Eq. \eqref{eq:StatWeightMulti} concerns only the types of anyons that exist as physical excitations in the system. Nevertheless, in a system where the anyons are so close that the fusion of anyons cannot be ignored, Eqs. \eqref{eq:StatWeight1} and \eqref{eq:StatWeightMulti} no longer hold. 

This does not mean that the ideal gas picture ceases being valid. In fact, the ideal gas picture may still apply and the thermodynamics of anyons can be accordingly derived, provided that the original state counting by Haldane and Wu is appropriately modified, even when there is just one species, as the fusion interaction of anyons can lead to a more generalized version of Eq. \eqref{eq:StatWeight1}. Consider for example $N$ doubled Fibonacci anyons with fusion interaction on a sphere. To handle this situation, In Ref.\cite{Hu2013}, the authors adopted the Levin-Wen model of topological orders, defined on a lattice with $P$ plaquettes on a sphere $S^2$, and introduced a new contribution to the statistical weight

\be\label{eq:StatWeightMultiPseudoSpices}
\begin{aligned}
        W_{\{G_i,N_i\}}=&\prod_{i=0}^{n-1}\bpm
        G_i+N_i-1-\sum_{j=0}^{n-1}\tilde\alpha_{ij}(N_j-\delta_{ij})\\
        \\
        N_i
        \epm
        \\&\qquad\times
        \prod_{i'=0'}^{n'-1}
        \sum_{N_{i'}=0}^{\Gamma}
        \bpm
        \tilde{G}_{i'}+N_{i'}-1-
        {\displaystyle \sum_{j=0}^{n-1}}\tilde\alpha_{{i'}j}(N_j-\delta_{{i'}j})
        -{\displaystyle \sum_{j'=0'}^{n'-1}}\tilde\alpha_{{i'}j'}(N_{j'}-\delta_{{i'}{j'}})       
        \\
        \\
        N_{i'}
        \epm,
\end{aligned}
\ee
where $i=0,\dots,n-1$ denotes the (real) anyon types. The second product factor is a new contribution, where $\Gamma=\lfloor\frac{1}{2}(\tilde{G}_{i'}-1-{\sum_{j=0}^{n-1}}\tilde\alpha_{{i'}j}(N_j-\delta_{{i'}j})-{\sum_{j'=0'}^{n'-1}}\tilde\alpha_{{i'}j'}(N_{j'}-\delta_{{i'}{j'}}))\rfloor$, and$i'=0',\dots,n'-1$ denotes the \textit{pseudo-species} \cite{Guruswamy1999a,Hu2013}. Pseudo-species anyons do not contribute to energy but to state counting via the $\tilde\alpha$ matrices. Hereafter, we use the primed indices to index the quantities associated to pseudo-species. Note that we denote the matrix of statistics parameters now by $\tilde\alpha$ instead of $\alpha$ as in Eq. \eqref{eq:StatWeightMulti}, to indicate the difference due to pseudo-species. 

The two summation terms in the second product shows that the number of possible ways to insert an additional pseudo-species anyon depends on the number of not only real anyons but also pseudo-ayons. The statistics is now characterized by the statistics parameters $\{\tilde{G},\tilde{\alpha}\}$, where $\tilde{G}_{i'}$ is the number of the single-particle vacancies for  the $i'$-th pseudo-species.

A doubled Fibonacci anyon $\tau\bar\tau$ appears as a fluxon excitation in a plaquette, and each at most one fluxon can inhabit a plaquette. By counting the number of states when there are $N$ fluxons allocated to $P$ plaquettes (as such, $P$ plays the role of $G$ in Eq. \eqref{eq:StatWeight1}), Ref.\cite{Hu2013} finds the following statistical weight.
\be\label{eq:StatWeightS2}
W^{S^2}_{P,N}=\bpm P\\N \epm
\sum_{N_1,N_2=0}^{\lfloor\frac{1}{2}(N-2)\rfloor}
\bpm  N-N_1-2\\ N_1 \epm
\bpm N-N_2-2\\ N_2\epm.
\ee 
Here, the first binomial factor $(\begin{smallmatrix}P\\N \end{smallmatrix})$ in fact corresponds to the right hand side (RHS) of Eq. \eqref{eq:StatWeight1}, while the other two binomial factors are new and due to the fusion of $\tau\bar\tau$. To mimic the RHS of Eq. \eqref{eq:StatWeightMulti}, The new binomial factors in Eq. \eqref{eq:StatWeightS2} lead to the following matrix of exclusion statistics.
\be\label{eq:alphatildeS2}
\tilde\alpha^{S^2}=\bpm
1 & 0 & 0\\
-1 & 2 & 0\\
-1 & 0 & 2
\epm.
\ee
This matrix is $3\x 3$ because in Eq. \eqref{eq:StatWeightS2}. There are just one type of the real anyons, i.e., the $N$ physical anyons $\tau\bar\tau$. There are also $N_1$ and $N_2$ anyons of two `hidden' species. These hidden species anyons do not appear as free excitations in the system and are named \textit{pseudo-species}\cite{Hu2013}.  A matrix element $\tilde\alpha_{ij}$ characterizes the statistical interaction between species (and/or types) $i$ and $j$, where $i=0$ labels the real anyons $\tau\bar\tau$, and $i=1,2$ label the two pseudo-species. More precisely, $\tilde\alpha_{ij}=n$ indicates that adding one more $j$-th species (or type) anyon, creates (if $n<0$) or kills (if $n>0$) $n$ single-particle vacancies for an additional $i$-th species (or type) anyon.

Note that the matrix in Eq. \eqref{eq:alphatildeS2} is denoted by $\alpha$ in Ref.\cite{Hu2013}; however, to emphasize its difference with the matrix $\alpha$ in Eq. \eqref{eq:StatWeightMulti}, we denote it by $\tilde\alpha$ as in Eq. \eqref{eq:StatWeightMultiPseudoSpices} and shall stick to this notation hereafter.

Two important questions remain unanswered, however, in Ref.\cite{Hu2013}. Firstly, what are the pseudo-species? Secondly,  how would gapped boundaries affect the state-counting and hence the exclusion statistics of anyons? We shall try to answer first the second question by considering anyon systems on open surfaces with explicit gapped boundary conditions. It will turn out that our answer to the second question help answer the first question too.

\subsection{Extended exclusion statistics on a multi-boundary open surface}

\begin{figure}[!h]
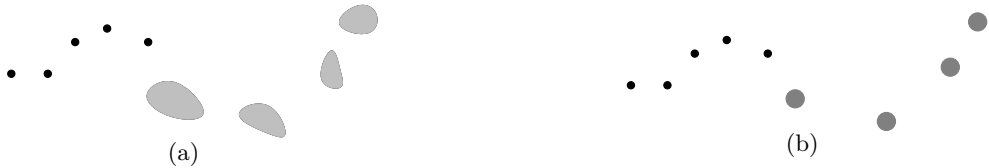

        \centering
        \subfigure[]{\AnyonAndBdries}
        \hspace{3cm}
        \subfigure[]{\AnyonAndBdriesAA} 
        \caption{(a) A system of 5 real elementary anyons and 4 gapped boundaries. Each gapped boundary can be transformed to a small hole as in (b), without affecting the Hilbert space structure of the system. In this perspective, the Hilbert space consists of 5 elementary anyons and 4 composite anyons respectively identified as the 4 boundaries components with their topological charges.}
        \label{fig:AnyonAndBdries}
\end{figure}

As argued earlier, previous studies of anyonic exclusion statistics had been restricted to closed spaces, i.e., without physical boundaries, but real-life materials usually have boundaries. This urges us to study the exclusion statistics of anyons on surfaces with gapped boundaries. See Fig. \ref{fig:AnyonAndBdries}(a) for an illustration. Please be warned of a bit terminology abusing: we use boundaries or boundary components interchangeably. 

For simplicity, in this paper we assume that all nontrivial topology is due to boundaries. As to be seen, however, our methods and results apply to generic Riemann surfaces with boundaries. 

Gapped boundary components can be deformed and shrunk to small holes, because of the topological properties of topological field theories (see Fig. \ref{fig:AnyonAndBdries}). Note here although we consider the boundary component disks to be topologically a point or small hole, we are assuming the size of the hole is large compared to the lattice locality scale such that the superposition of charge types in the hole does not spontaneously decohere. The relevant observables of these small disks are respectively the total topological charges of the disks. Hence, to characterize the Hilbert space, we treat each boundary component as a composite anyon (a direct sum of the elementary anyons in the system). The decomposition of such composite anyons into elementary anyon species is determined by the corresponding boundary types. In this perspective, the state counting on an $M$-boundary surface amounts to that on a sphere with $M$ extra composite anyons. The elementary anyons and the composite anyons together obey the exclusion statistics governed by Eq. \eqref{eq:StatWeightMulti}, if fusion interaction is ignored, and \eqref{eq:StatWeightMultiPseudoSpices} otherwise.

The statistical weight then depends on the number and types of boundary components, in the same way as how it depends on the number and species of real elementary anyons present in the system.  

This perspective enables us to extend the exotic exclusion statistics to anyon systems on open surfaces: \textit{The number of effective single particle states mutually depends on the number of (all species) existing anyons as well as the number of (all types of) boundary components.} We shall christen the exotic exclusion statistics of anyons in such cases \textit{extended anyonic exclusion statistics}.

The dependence relation can be expressed by the linear equation
\begin{equation}\label{eq:linearEqExtended}
\Delta{d_i}=-\alpha_{ij}\Delta{N_j},
\end{equation}
in the same form as originally proposed by Haldane\cite{Haldane1991a}. Here $\Delta{N_j}$ is the allowed change of the total particle number for species $j$, and $d_i$ is the number of available single particle states for species $i$. $d_j$ is generally not a constant but depends on $\{N_j\}$. The above equation states that, when adding an additional $j$-particle, the number of available single particle states for all species $i$ changes as $\Delta{d_i}=-\alpha_{ij}$. The statistics parameter $\alpha_{ij}$ defines the mutual statistics between anyons.

We extend the relation to a gapped system with boundaries in the following way. We keep the same form of the equation, while interpreting $N_i$ as the particle number for particle species $i$ when $0\leq i \leq n-1$, but as the number of boundary components for boundary type $i$ when $n\leq i \leq n+m-1$. where $m$ and $n$ are respectively the total number of boundary types and that of real anyon species in the system. Now The statistics parameter $\alpha_{ij}$ defines the mutual statistics among anyons and boundary components.

The extended exclusion statistics quantitatively describe the bulk-boundary correspondence. The statistics parameters $\{\alpha_{ij}\}$ specify the linear difference relation how the number of effective single particle states for $i$-species depends on the number of existing species-$j$ anyons (for $0\leq j\leq n-1$), as well as on the type-$j$ boundaries (for $n\leq j \leq n+m-1$). The formula above explicitly shows that the two factors are on an equal footing.

Because any number and species of elementary Abelian anyons always fuse to a definite elementary Abelian anyon, the notion of pseudo-species is unnecessary even if fusion is considered. Hence, the extended statistical weight is given by
\begin{equation}\label{eq:StatWeightWithBdry}
\small
W_{\{N_i,M_k\}}=\prod_{i=0}^{n+m-1}\bpm
G_i+N_i-1-\sum_{j=0}^{n+m-1}\alpha_{ij}(N_j-\delta_{ij})\\
\\
N_i
\epm
\end{equation}

For a non-Abelian anyon system, according to Ref. \cite{Hu2013}, we need to bring in pseudo-species anyons. We propose the general formula of the extended statistical weight:

\be\label{eq:ReducedStatWeightWithBdry}
\small
\begin{aligned}
        W_{\{G_i,N_i\}}=&\prod_{i=0}^{n+m-1}\bpm
        G_i+N_i-1-\sum_{j=0}^{n+m-1}\tilde\alpha_{ij}(N_j-\delta_{ij})\\
        \\
        N_i
        \epm
        \\&\qquad\times
        \prod_{i'=0'}^{n'+m'-1}
        \sum_{N_{i'}=0}^{\Gamma'}
        \bpm
        \tilde{G}_{i'}+N_{i'}-1-
        {\displaystyle \sum_{j=0}^{n+m-1}}\tilde\alpha_{{i'}j}(N_j-\delta_{{i'}j})
        -{\displaystyle \sum_{j'=0'}^{n'+m'-1}}\tilde\alpha_{{i'}j'}(N_{j'}-\delta_{{i'}{j'}})    
        \\
        \\
        N_{i'}
        \epm,
\end{aligned}
\ee
where $\Gamma'=\lfloor\frac{1}{2}\tilde{G}_{i'}-1-{\sum_{j=0}^{n+m-1}}\tilde\alpha_{{i'}j}(N_j-\delta_{{i'}j})-{\sum_{j'=0'}^{n'+m'-1}}\tilde\alpha_{{i'}j'}(N_{j'}-\delta_{{i'}{j'}})\rfloor$, and $n'$ and $m'$ denotes the number of pseudo-species associated with the real anyons and the boundaries, respectively. The meaning of these pseudo-species will be clear later. The statistics is characterized by the statistics parameters $\{\tilde{G},\tilde{\alpha}\}$. Eq. (\ref{eq:ReducedStatWeightWithBdry}) also holds for Abelian anyons on open surface, with $n'$ part being trivial and all $\tilde\alpha_{i'j}$, $\tilde\alpha_{i'j'}$ being zero except for the $\tilde\alpha_{i'i'}=1$.

For future convenience, let us name the factors due to pseudo-species in the statistical weight \eqref{eq:ReducedStatWeightWithBdry} the \textit{reduced statistical weight} $w_{\{N_i\}}$, i.e.,
\begin{equation}\label{eq:ReducedStatisticaWeight}
\begin{aligned}
w_{\{N_i\}}=
\prod_{i'=0'}^{n'+m'-1}
\sum_{N_{i'}=0}^{\Gamma'}
\bpm
\tilde{G}_{i'}+N_{i'}-1-
{\displaystyle \sum_{j=0}^{n+m-1}}\tilde\alpha_{{i'}j}(N_j-\delta_{{i'}j})
-{\displaystyle \sum_{j'=0'}^{n'+m'-1}}\tilde\alpha_{{i'}j'}(N_{j'}-\delta_{{i'}{j'}})    
\\
\\
N_{i'}
\epm,
\end{aligned}
\end{equation}

\section{Anyonic exclusion statistics on a disk}\label{sec:Disk}
In this section, we study the extended anyonic exclusion statistics on a disk with a gapped boundary. We shall show that the extended statistical weight indeed satisfies (in fact a special case of) Eq. \eqref{eq:ReducedStatWeightWithBdry}. We shall reveal the true meaning of pseudo-species too.
The next section will generalize the discussion and results here to the case of multiple boundaries.
 
In principle, there are three methods of state-counting: 1) model-based counting, 2) counting by state basis, and 3) counting by fusion channels. The model-based method is unambiguous, as the calculation is based on well-established models of topological orders. Having obtained trustworthy results using the model-based method, and corroborated by the counting by fusion space, we will know how we should construct the correct bases of the relevant Hilbert spaces.
It will turn out that method 2) counting by state basis is the most convenient and systematic method to extract the extended statistical weight and deepen our understanding of pseudo-species.
\subsection{The extended Levin-Wen model and the fluxon-number operators}
\label{subsec:eLWfluxon}
\begin{figure}[!h]
        \centering
        \includegraphics[scale=0.3]{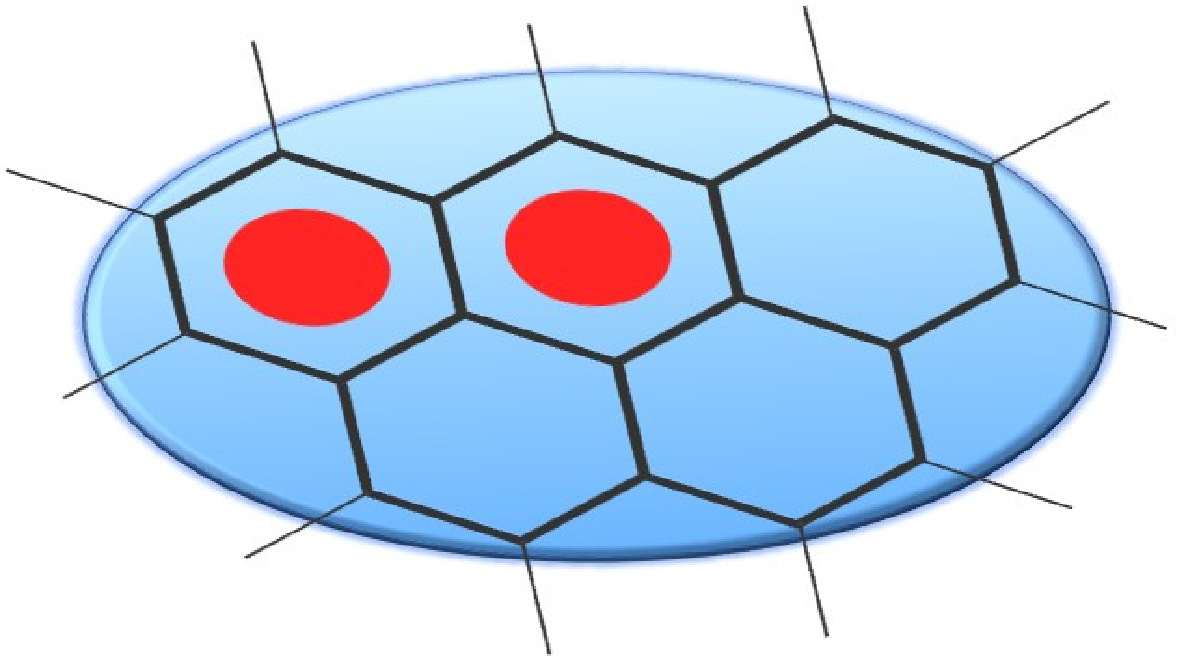}
\caption{The extended Levin-Wen model defined on a hexagonal lattice embedded in a disk. Thick (thin) black lines are the bulk (boundary) edges of the lattice. The two red dots represent two $\tau\bar\tau$'s on the two plaquettes, for instance.}
\label{fig:LWdisk}
\end{figure}

The model we take is the extended Levin-Wen model with boundaries, developed by two of us\cite{Hu2017,Hu2017a}, based on the original Levin-Wen model\cite{Levin2004}. Alternative work can be found in \cite{Kitaev2012}.The model is an exactly solvable Hamiltonian model defined on a trivalent lattice. The model's topological properties are invariant under lattice mutations that preserve the topology; hence, any planar trivalent graph works, and lattice size does not matter. For simplicity and a common choice, we take hexagonal lattices. A hexagonal lattice embedded on a disk is shown in Fig. \ref{fig:LWdisk}. The degrees of freedom all live on the edges of the lattice. The bulk degrees of freedom take value in a unitary fusion category (UFC), namely a finite set of labels $L=\{s=a,b,c,\dots\}$ that satisfy a set of fusion rules
\be \label{eq:fusion}
a\ox b=\sum_c N^{c}_{ab}c,
\ee
$\forall a,b,c\in L$, where $N^c_{ab}\in\Z_{\geq 0}$ are the fusion coefficients.
Each object $s\in\ L$ is called a string type that comes with a defining quantum number $d_s$, called the quantum dimension. The total quantum dimension $D$ is defined by $D=\sum_s d_s^2$.
The degrees of freedom living on the boundary edges take value in a subset $A\subseteq L$. The fusion rules of the objects in $A$ inherit those of $L$ by removing any objects not in $A$ from the RHS of Eq. \eqref{eq:fusion}.
By definition, $D_A=\sum_{s\in A}d_s$. The subset $A$ under the fusion rules forms a Frobenius algebra object of the UFC $L$\cite{Hu2017,Hu2017a}. The Hamiltonian of the model reads
\be
H=H_{\mathrm{bulk}}+H_{\mathrm{bdry}},
\ee
with
\be\label{eq:bulkHam}
H_\mathrm{bulk}=-\sum_{v\in \mathrm{bulk}} A_v -\sum_{p\in \mathrm{bulk}} \frac{1}{D}\sum_{s\in L}d_s B_p^s,
\ee 
and
\be\label{eq:bdryHam}
H_\mathrm{bdry}=-\sum_{v\in \mathrm{bdry}} \bar A_v -\sum_{p\in \mathrm{bdry}} \frac{1}{D_A}\sum_{s\in A}d_s \bar B_p^s,
\ee
where the sums run over the vertices and plaquettes of the lattice respectively. Appendix \ref{appd:LW} records more details of the operators in the above Hamiltonians. 

In this model, the bulk anyon excitations form a unitary modular tensor category (UMTC) $\C=L\x \bar L$, where $\bar L=\{\bar s=0,\bar 1,\bar 2,\dots,\bar m \}$ is the time reversal of $L$. Same fusion rules apply to $\bar L$.  A fusion rule of two anyons in $\C$ is simply the product of the corresponding fusion rules in $L$ and $\bar L$, e.g. $a\bar b\ox c\bar d=(a\ox c)(\bar b\ox \bar d)$. The quantum dimension of any $a\bar b\in\C$ is $d_{a\bar b}=d_a d_{\bar b}=d_a d_b$. In particular, the diagonal pairs $a\bar a\in C$ are called the fluxons who live only in the plaquettes of the lattice, and at most one fluxon can inhabit each plaquette. Other excitations are either chargeons living at the vertices or dyons as chargeon-fluxon bound states. Therefore, the original Levin-Wen model is said to effectively describe a doubled topological order.

Excitations on a boundary, however, correspond to the bulk anyons that can condense at the boundary and are responsible for gapping the boundary\cite{Levin2012,Kong2013,HungWan2013a,HungWan2014}. Condensable anyons are dictated by the boundary Hamiltonian \eqref{eq:bdryHam}. It turns out that in such a doubled topological order, all diagonal pairs $a\bar a$---the fluxons---can condense at a boundary, rendering the boundary fully gapped.

Now let us focus on a specific case where $L_{\text{Fib}}=\{1,\tau\}$ with fusion rules $1\ox\tau = \tau$ and $\tau \ox \tau =1+\tau$. The quantum dimensions are $d_1=1$ and $d_\tau=\frac{\sqrt{5}+1}{2}=:\phi$. This UFC $L_{\text{Fib}}$ is called the Fibonacci category. With $L_{\text{Fib}}$ being the input data, the model output the doubled Fibonacci topological order $\C_{\text{Fib}}=\{1,\tau 1,1\bar\tau,\tau\bar\tau\}$ with obvious fusion rules and quantum dimensions. The trivial anyon $1$ and $\tau\bar\tau$ are the two fluxons in this case. We want to see for a given number $P$ of bulk plaquettes, a given number of $N$ fluxons $\tau\bar\tau$ in the bulk, and the gapped boundary condition due to $\tau\bar\tau$ condensation at the boundary, what is the number of states? 

To do the state counting to answer the question, we can construct the fluxon-number operators that count the numbers of fluxons in the bulk and at the boundary. The bulk fluxon-number operators $n^1_p$ and $n^{\tau\bar\tau}_p$ for a bulk plaquette $p$ are defined by
\be\label{eq:fluxNumOp}
\begin{aligned}
& n^1_p=\frac{1}{D}(B^1_p+\phi B^\tau_p),\\
& n^{\tau\bar\tau}_p=\frac{1}{D}(\phi^2 B^1_p-\phi B^\tau_p),
\end{aligned}
\ee
 where $B^1_p\equiv\mathds{1}$ and $B^\tau_p$ are the bulk plaquette operators in the bulk Hamiltonian \eqref{eq:bulkHam}. All fluxon number operators are projectors, and commute with one another and with all $A_v$ and $B_p$ operators. If $n^{\tau\bar\tau}_p=1$, there is a $\tau\bar\tau$ excitation in the plaquette $p$. The same applies to $n^1_p$; however, since $1$ is the trivial anyon or vacuum, $n^1_p=1$ means no nontrivial anyon in plaquette $p$. Note also that $n^1_p n^{\tau\bar\tau}_p=0$ and $n^1_p+n^{\tau\bar\tau}_p=1$ hold in the $A_v=1$ subspace to which we restrict our computation; hence, they are orthonormal projectors. According to Ref.\cite{HungWan2014,Hu2017a}, the doubled Fibonacci topological order has two morita-equivalent Frobenius algebras that doesn't affect the calculation, hence it has only one gapped boundary condition, which is due to the $\tau\bar\tau$ condensation. That is, the fluxon number operators in Eq. \eqref{eq:fluxNumOp} have their boundary versions
\be\label{eq:fluxNumOpBdry}
\begin{aligned}
&\bar n^1_p=\frac{1}{D_A}(\bar B^1_p + \sqrt{\phi} \bar B^\tau_p),\\
&\bar n^\tau_p=\frac{1}{D_A}(\phi \bar B^1_p-\sqrt{\phi}\bar B^\tau_p),
\end{aligned}
\ee
where $\bar B^1_p\equiv \mathds{1}$ and $\bar B^\tau_p$ are the boundary plaquette operators in Eq. \eqref{eq:bdryHam}. 

\subsection{Model-based state counting}
Our goal is to count the number of states when $\Ntbt$ fluxons $\tau\bar\tau$ exist in the bulk of a lattice $\Gamma$ with $P$ bulk plaquettes and no any other excitations in the bulk or boundary at all. Since the fluxons must live in exactly $\Ntbt$ bulk plaquettes, there are $(\begin{smallmatrix}P \\N_{\tau\bar\tau}\end{smallmatrix})$ configurations of taking $\Ntbt$ out of $P$ plaquettes. In each configuration $C$ of fixed $N_{\tau\bar\tau}$ plaquettes, each plaquette indwells a $\tau\bar\tau$, and all the rest $P-\Ntbt$ plaquettes not in $C$ do not indwell any nontrivial anyon. Therefore, the total statistical weight $W^{D^1}_{P,N}$, i.e., the total number of states , would be the sum of the reduced statistical weights $w_{P,\Ntbt,C}$ respectively in each configuration $C$. Namely,
\be\label{eq:Wdisk}
\begin{aligned}
W^{D^1}_{P,\Ntbt}=\bpm P\\\Ntbt\epm w_{\Ntbt,C}=\bpm P\\ \Ntbt \epm \mathrm{Tr}\left(\prod_{p\in C} n^{\tau\bar\tau}_p
\prod_{p\notin C} n^1_p \prod_{p\in\partial\Gamma} \bar n^1_p \right),
\end{aligned}
\ee
where $\partial\Gamma$ denotes the boundary of the lattice $\Gamma$, and the trace runs over the entire Hilbert space. Here we denote a disk by $D^1$. Note again that the invariance of the topological properties of the model under topology-preserving lattice mutations renders the statistical weight independent of the lattice being considered, as long as $P$ does not change. If we focus on the reduced statistical weight $w_{\Ntbt,C}$, $P$ becomes irrelevant as long as $P\geq N_{\tau\bar\tau}$. Hence, we may just pick any trivalent lattice $\Gamma$ with $P$ plaquettes convenient for our computation. 

We compute using Eq. \eqref{eq:Wdisk} and the precisely the lattice in Fig. \ref{fig:LWdisk}, namely with $P=5$, and obtain the state counting in Table \ref{tab:diskFibW}. If we increase the plaquette number $P$, we can obtain the state counting for larger $\Ntbt$, which are neglected into the `$\cdots$' in the table.
\begin{table}[!h]
\caption{State counting of $\tau\bar\tau$ on a disk.}\label{tab:diskFibW}
\centering
\begin{tabular}{|m{4em}|m{2em}m{2em}m{2em}m{2em}m{2em}m{2em}m{2em}|}
\hline 
$\Ntbt$ & $0$ & $1$ & $2$ & $3$ & $4$ & $5$ & $\cdots$ \\  \hline 
$w_{P,N,C}$ & $1$ & $1$ & $2$ & $5$ & $13$ & $34$ & $\cdots$ \\ 
\hline
\end{tabular}
\end{table}

We can then fit the data in Table \ref{tab:diskFibW} by binomials and cast Eq. \eqref{eq:Wdisk} in the form
\be\label{eq:WdiskBin}
W^{D^1}_{P,\Ntbt} =\bpm P\\\Ntbt\epm 
\sum_{N_1=0}^{\Ntbt-1} \bpm
2\Ntbt-N_1-2 \\ N_1
\epm,
\ee
where $N_1$ counts the anyons of the pseudo-species, and there is only one pseudo-species on a disk. This statistical weight is clearly a special case of Eq. \eqref{eq:ReducedStatWeightWithBdry} when there is only one type of real anyons, one type of boundaries, the number of boundaries is one, and only one pseudo-species. Hence, in light of Eqs. \eqref{eq:StatWeightMulti}-\eqref{eq:alphatildeS2}, the second binomial, i.e., the reduced statistical weight, in the statistical weight above leads to the following matrix of exclusion statistics.
\be\label{eq:alphaDisk}
\atildedisk=\bordermatrix{
  & 0 & 1\cr
0 & 1 & 0\cr
1 & -2 & 2 },
\ee
where we explicitly placed the row and column indices. 

Let us first briefly explain the entries of the matrix above. The index $i=0$ labels the real anyons, i.e., the fluxons $\tau\bar\tau$, as physical excitations in the bulk of the disk.
The index $i=1$ labels the pseudo-species, whose meaning will be elaborated later. With this labeling, from the reduced statistical weight in \eqref{eq:WdiskBin}, we can also extract the single-particle state number $\tilde G_1=-3$ for the pseudo-species. The more useful quantity is actually the effective single-particle state number $\tilde G_{1}^{eff}$ (similar to that in Eq. \eqref{eq:StatWeight1}) defined as
\be\label{eq:G1effDisk}
\begin{aligned}
\tilde G_1^{eff} & = \tilde G_1 -\sum_{j=0}^1 (\atildedisk)_{1j}(N_j-\delta_{1j})\\
& = \tilde G_1 -(\atildedisk)_{10}N_0 - (\atildedisk)_{11}(N_1-1)\\
& = \tilde G_1 -(\atildedisk)_{10}N_0 \\
& = -3 + 2 N_0,
\end{aligned}
\ee
where the third equality is obtained by letting $N_1=1$ because we are concerned about merely the single-particle state vacancies of the pseudo-species. From $\tilde G_1^{eff}(N_0) = -3 + 2 N_0$, we have $\tilde G_1^{eff}(2)=1$, i.e. there should be at least two real fluxons $\tau\bar\tau$ for the pseudo-species to have one single-particle vacancy.  

Moreover, $\tilde G_1^{eff}(N_0+1) - \tilde G_1^{eff}(N_0) \equiv  (\atildedisk)_{10} = 2$, i.e., adding one more real $\tau\bar\tau$ increases the single-particle vacancies of the pseudo-species by $2$. Actually, in general, the entry $(\atildedisk)_{ij}$ indicates that for each $j$-anyon added into the system, there would be $(\atildedisk)_{ij}$ vacancies fewer for the $i$-anyon single-particle states if $(\atildedisk)_{ij}>0$, or more if $(\atildedisk)_{ij}<0$. If $(\atildedisk)_{ij}=0$, the $j$-anyons do not affect the vacancies for $i$-anyons. In Eq. \eqref{eq:alphaDisk}, $(\atildedisk)_{00}=1$ is consistent with that each fluxon $\tau\bar\tau$ can occupy at most one plaquette of the lattice. The entry $(\atildedisk)_{11}=2$ indicates the following exclusion rule for the pseudo-species anyons.
\begin{table}[!h]
\centering
\begin{tabular}{|C|C|C|C|C|}
\hline
$\checkmark$ & $\x$ & $\blacksquare$ & $\x$ & $\checkmark$ \\
\hline
\end{tabular}
\end{table}

In the exclusion rule above, we define a box as a vacancy and call it a vacancy box. A $\boxed{\blacksquare}$ is a vacancy box occupied, a $\boxed{\x}$ is a vacancy box that cannot be occupied (or unhabitable), and $\boxed{\checkmark}$ is a vacancy box that may be occupied (or habitable). In other words, for the pseudo-species, if a vacancy box is occupied, the two neighbouring vacancy boxes become unhabitable. A vacancy box is analogous to an orbital of an electron in an atom, say, for example. 

One can see that the exclusion rule above is already beyond those for bosons and fermions, as a vacancy box occupied by a boson or fermion does not forbid a neighboring vacancy box to be occupied because $\alpha_{\text{boson}}=0$ and $\alpha_{\text{fermion}}=1$, as pointed out in Section \ref{sec:rev}. We are yet able to precisely identify what the pseudo-species anyons are, however, because the rule above is insufficient for us to derive the fusion rules of the pseudo-species anyons. To obtain a complete set of exclusion rules, we need to understand more deeply about the entry $(\atildedisk)_{10}=-2$. It will turn out that the simple exclusion rule above needs to be revised, in the sense that a single vacancy box is not enough for specifying the single-particle states of pseudo-species. The entry $(\atildedisk)_{10}=-2$ indicates that adding a real fluxon $\tau\bar\tau$ would add two more vacancy boxes for the pseudo-species. The question is: how are the two vacancies added?

\section{Systematic basis construction on a disk}\label{sec:basisDisk}
To answer the question above, it is better to build a basis of the multi-anyon state space. The total Hilbert space is a tensor product of two spaces. The first is the space of configurations of $N$ anyons in the system, such as arranging $N$ fluxons $\tau\bar\tau$ among $P$ plaquettes in the extended Levin-Wen model. The second is the fusion Hilbert space results from the fusion of the anyons. The fusion Hilbert space is the only space relevant to the reduced statistical weight. From now on, we shall focus on the reduced statistical weight and call the fusion Hilbert space simply by Hilbert space for short. 
\subsection{Constructing a generic basis}
We first look at the Hilbert space of two to five fluxons $\tau\bar\tau$.
\begin{align*}
\tau \bar \tau \ox \tau \bar \tau & = 1 +\tau \bar \tau+ 1\bar\tau + \tau1 ,
\\
(\tau \bar \tau)^3 &=1 +4 \tau \bar \tau+ 2(1\bar\tau+  \tau1),
\\
(\tau \bar \tau)^4 &= 4(1  +\tau \bar \tau) +5\tau \bar \tau + 6(\tau1 +1\bar \tau ),\\
(\tau \bar \tau)^5 &=9 (1  +\tau \bar \tau)+16 \tau\bar\tau + 15(\tau \bar1 + 1\bar\tau ),\\
&\ \, \vdots
\end{align*}
Comparing the equations above with Table \ref{tab:diskFibW}, we find that the state counting in Table \ref{tab:diskFibW} actually counts the number of fusion channels to trivial fluxons $1$ and fluxons $\tau\bar\tau$ appearing on the RHS of each of the equations above. If we did the state counting in this way, we would be using the method of counting by fusion channels, mentioned in the beginning of Section \ref{sec:Disk}. Nonetheless, in general cases with many different boundary conditions, it is not easy to decide which fusion channels we should select.\footnote{\label{fn:multiplicity}According to Ref. \cite{HungWan2014}, anyons condensing at a boundary can appear with multiplicities, such that they contribute to ground states multiple times.} Hence, it is better to find a more systematic method.

\begin{figure}[!h]
\centering
\includegraphics[scale=1.2]{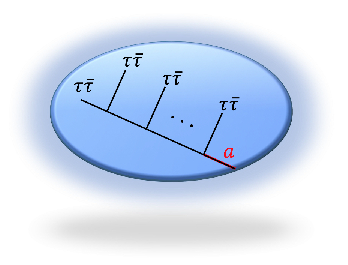}
\caption{(Color online.) A sketch of the basis states of multi-$\tau\bar\tau$ Hilbert space on a disk. The line $a$ in red represents either $1$ or $\tau\bar\tau$.}
\label{fig:diskBasis}
\end{figure}

The state counting using the extended Levin-Wen model therefore tells us which subspace of a multi-fluxon Hilbert space should be singled out as the physical Hilbert space. This result complies with that the gapped boundary of the disk is due to condensing $\tau\bar\tau$ at the boundary. On a sphere, the Hibert space of multi-$\tau\bar\tau$ consists of the fusion channels of these $\tau\bar\tau$ into the trivial anyon $1$. That is, the total topological charge of the system must be trivial. On a disk, however, according to Ref.\cite{HungWan2014}, when a $\tau\bar\tau$ condenses at the boundary, it  also contributes a copy of the trivial anyon $1$. Thus, the physical Hilbert space of multi-$\tau\bar\tau$ on a disk consists of the fusion channels of these $\tau\bar\tau$ into either $1$ or $\tau\bar\tau$; the final fusion product $\tau\bar\tau$ is the total charge of the real $\tau\bar\tau$'s and must be annihilated at the boundary, as otherwise there would be excessive fluxons in the system. This discourse motivates the basis in Fig. \ref{fig:diskBasis}. Since, fusion is associative, we can always present the fusion of multiple fluxons $\tau\bar\tau$ as a tree graph as in Fig. \ref{fig:diskBasis}. 

In Fig. \ref{fig:diskBasis}, leaves represent the real fluxons $\tau\bar\tau$; the root (red edge) $a$ of the tree is the fusion product of all $\tau\bar\tau$'s in the bulk and can be only either $1$ or $\tau\bar\tau$; the trunk edges (black internal edges along the trunk) can take value in $\{1, \tau1, 1\tau, \tau\bar\tau\}$. The physical Hilbert space is spanned by all possible such fusion trees. Later, we shall see that the trunk edges are actually place-holders for pseudo-species anyons. Moreover, the root will turn out to be a place-holder for another pseudo-species in cases with more than one boundary. The boundary itself then can be viewed as a composite anyon $1\oplus\tau\bar\tau$,\footnote{This viewpoint does not suffer the issue in footnote \ref{fn:multiplicity} caused by the multiplicity of an anyon condensing at the boundary. Suppose an anyon $a$ condensing at a boundary has multiplicity $2$, we would simply denote the boundary as composite anyon $1\oplus 2a$ or $1\oplus a\oplus a$.} as alluded to in Section \ref{sec:rev}. In this perspective, the total charge of the entire system is still trivial.

To understand this claim, let us turn the basis graph in Fig. \ref{fig:diskBasis} more schematic and abstract, and with the help of $\atildedisk$, we can decipher the pseudo-species. 
\begin{figure}[!h]
\centering
\includegraphics[scale=1]{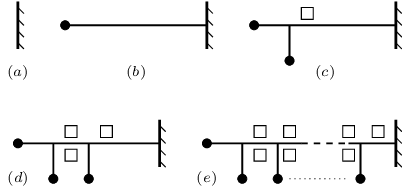}
\caption{Our convention for the basis of multi-$\tau\bar\tau$ Hilbert space on a disk. A $\tau\bar\tau$ in the bulk is denoted by a black dot. A place-holder for the pseudo-species is a box. (a) Abstract presentation of the disk's boundary. We always draw the boundary on the right.  (b) A single $\tau\bar\tau$ in the bulk; it must be attached to the boundary by an edge. (c) Adding the second $\tau\bar\tau$, which introduces a place-holder on the root of the tree. We always add new $\tau\bar\tau$'s from the left to right. (d) Adding the third $\tau\bar\tau$, which introduces a place-holder on its left (the lower box) and one on its right (the upper box). (e) The basis with place-holders for multiple $\tau\bar\tau$'s. The root always has an upper box.}
\label{fig:basisDiskWithBox}
\end{figure}

We denote the boundary of a disk as in Fig. \ref{fig:basisDiskWithBox}(a), and we always present the boundary on the right, as when we track along the boundary, the bulk is on the left. Note again that in this paper, each boundary is compact and closed, i.e., an $S^1$. As discussed earlier, on a disk, a single $\tau\bar\tau$ can exist in the bulk but has to be attached to the boundary by a string\footnote{The string can be thought as a string operator in the extended Levin-Wen model or a Wilson string in the effective topological field theory describing the topological order.}, as in Fig. \ref{fig:basisDiskWithBox}(b), where the $\tau\bar\tau$ is represented by a black dot. The basis tree has merely a single leaf.  

We can then add one $\tau\bar\tau$ each time into the bulk. Adding the second $\tau\bar\tau$ would lead to the basis in Fig. \ref{fig:basisDiskWithBox}(c), where the vertex is due to the fusion of two $\tau\bar\tau$'s. Now the basis tree has a root and two leaves. The root represents an internal state that is either $1$ or $\tau\bar\tau$ because the Hilbert space is doubly degenerate according to Table \ref{tab:diskFibW}. It is then natural to add a place-holder (the box in the figure) on the root to bookkeep this internal state. To get a better understanding, we invoke the $\tilde G_1^{eff}$ in Eq. \eqref{eq:G1effDisk} and the discussion below the equation. So as the pseudo-species to have a positive number of single-particle states, since $\tilde G_1^{eff}= -3 + 2N_0$, there must be at least two real $\tau\bar\tau$'s, i.e., $N_0\geq 2$. Now that $\tilde G_1^{eff}(N_0=2)=1$, adding the second $\tau\bar\tau$ to the disk would add one and only one place-holder, i.e., a vacancy box, for the pseudo-species. The pseudo-species must reside on the root---the only internal edge---of the tree in Fig. \ref{fig:basisDiskWithBox}(c) because all other edges are leaves representing real anyons. Now a place-holder is identified with a vacancy box just defined, and as our convention, we always place the box above the root. Since we know what the internal states can be, we can deem the empty box being $\tau\bar\tau$ while the occupied box $\boxed{\blacksquare}$ being $1$. We will get back to this assignment later.

Now let us add the third $\tau\bar\tau$ in to the bulk. We choose to add a $\tau\bar\tau$ always to the right of the existing ones and existing boxes, and below the trunk. This is merely a convention, as the associativity of fusion and that the boundary is $S^1$ allow us to move a leaf anywhere on the tree. We shall stick to this convention hereafter. Fig. \ref{fig:basisDiskWithBox}(d) depicts the $3$-$\tau\bar\tau$ basis. 

Here comes an important point. Pseudo-species arise to account for the exclusion statistics of anyons when the fusion of the anyons are taken into account. According to $(\atildedisk)_{10}=-2$ in Eq. \ref{eq:alphaDisk}, adding one real $\tau\bar\tau$ would increase the vacancy boxes of the pseudo-species by $2$, an artifact of the fusion. As a trunk edge (internal edge) in a basis tree is a result of a fusion, it is reasonable to put the vacancy boxes on each trunk edge. Staring at Figs. \ref{fig:basisDiskWithBox}(c) and (d), adding one more $\tau\bar\tau$ leads to one more junction that separates the root into a trunk edge and again the root. This turns the box place-holder on the root in Fig. \ref{fig:basisDiskWithBox}(c) into a box above the new trunk edge. Hence, as dictated by $(\atildedisk)_{10}=-2$, to add $2$ vacancy boxes for the pseudo-species, we choose to add one below the new trunk edge and the other one again above the root. Effectively, as in Fig. \ref{fig:basisDiskWithBox}(d), we still have only one box above the root, i.e., two boxes in total, for the new trunk edge, where a pseudo-species anyon resides. 

Following the recipe of adding more and more $\tau\bar\tau$'s into the disk, the basis of multi-$\tau\bar\tau$ Hilbert space, with explicit vacancy boxes for the pseudo-species, appear to be the one in Fig. \ref{fig:basisDiskWithBox}(e). There, each trunk edge has one box above and one box below, and the root has a single box above. 

\begin{figure}[!h]
\centering
\includegraphics[scale=1]{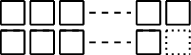}
\caption{Simplified basis of a multi-$\tau\bar\tau$ Hilbert space on a disk. the upper right box is the one above the root in Fig. \ref{fig:basisDiskWithBox}(e) and is accompanied with a dashed box underneath.}
\label{fig:boxBasisDisk}
\end{figure}

Now the question is: How we may decide which vacancy boxes in the disk basis Fig. \ref{fig:basisDiskWithBox}(e) can be occupied and which cannot be? To find the answer, let us further simplify the basis into the form in Fig. \ref{fig:boxBasisDisk} by removing the tree and the boundary but keeping the boxes only. This procedure is correct because the degrees of freedom of the Hilbert space being studied reside only in the vacancy boxes. In this simplified basis, a dashed box is added below the upper right box for reasons to be clear shortly. This new basis comprises of solely the pseudo-species vacancy boxes but keep in mind that between two neighbouring columns, there is a real $\tau\bar\tau$ inserted.

\subsection{Generalized Pauli exclusion principles}
Aided by the matrix $\atildedisk$ \eqref{eq:alphaDisk}, we can then determine the rules of how the boxes in the basis in Fig. \ref{fig:boxBasisDisk} should be occupied. The two boxes in each column are a result of $\atildedisk=-2$, namely, one extra real $\tau\bar\tau$ adds two more vacancy boxes for the single-particle state of the pseudo-species. That is, \textit{each configuration of occupying the two boxes in a column corresponds to exactly one single-particle state of the pseudo-species}. 

The right most column is the result of $\tilde G_1^{eff}=-3 + 2N_0 = 1|_{N_0=2}$, which is a boundary effect, as explained earlier. Since there is just one vacancy box for the single-particle state of the pseudo-species at the right end, we can always duplicate the box by adding a virtual box (the dashed box in Fig. \ref{fig:boxBasisDisk}) below it and deem the two boxes configured the same way.

Recall for a bosons, $\alpha=0$; hence, a vacancy box for bosons can be occupied arbitrary number of times. For a fermion, however, $\alpha=1$; hence, a vacancy box for fermions can be occupied at most once. Now we have $(\atildedisk)_{11}=2$; hence, not only a vacancy box of the pseudo-species can be occupied at most once but also the nearest neighbouring boxes in the same row cannot be occupied, as a generalization of the Pauli exclusion principle.  We have briefly touched upon such exclusion rules earlier. The new ingredient here is that each pseudo-species anyon takes up two vacancy boxes, i.e., the two boxes in a column in Fig. \ref{fig:boxBasisDisk}. Remember also that the two vacancy boxes are not added at the same time, their configurations are independent. Thus, when a box in a column is occupied, it cannot exclude the other box in the same column being occupied because it takes the entire column to specify a single-particle state of the pseudo-species. As such, an occupied box can only expel the occupation of its nearest neighbouring boxes. The right most column is an exception because there we put in a duplicate box by hand. 

A generalized Pauli exclusion principle on the disk can then be coined as follows.
\begin{principle}\label{pr:excluDisk}
\begin{equation*}
\includegraphics{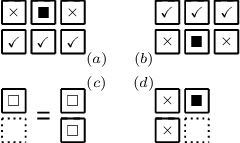}
\end{equation*}
Here, $\boxed{\square}$ is a wildcard, meaning the box is either occupied or unoccupied; the RHS of $(c)$ forces the two boxes in the column to be both occupied (unoccupied) if the box on the LHS is occupied (unoccupied).
\end{principle}

Exclusion Principle \ref{pr:excluDisk}(d) is actually redundant because it can be inferred from (a) and (c); however, to make the rules more explicit, as this is the first time we encounter such exclusion principles, we include (d). In general cases with multiple boundaries, such as a cylinder, we shall list only the minimal set of independent principles. 

\begin{figure}[!h]
\centering
\includegraphics[scale=1]{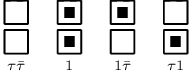}
\caption{Identifying the single-particle states of the pseudo-species, as a solution to Eq. \ref{eq:fusionEqDisk}.}
\label{fig:pseudoID}
\end{figure}

From Exclusion Principle \ref{pr:excluDisk}, we can deduce the single-particle states of the pseudo-species precisely and recover their fusion rules. To this end, note that in between two neighbouring columns in the basis in Fig. \ref{fig:boxBasisDisk}, there is a real $\tau\bar\tau$ (recall Fig. \ref{fig:basisDiskWithBox}(e)); hence, Exclusion Principle \ref{pr:excluDisk} leads to the following equations of fusion.  
\be\label{eq:fusionEqDisk}
\begin{aligned}
\boxone \otimes \tau\bar\tau & = \boxttbar, \\
\boxttbar \otimes \tau\bar\tau & = \boxone + \boxonetbar + \boxtone + \boxttbar, \\
\boxonetbar \otimes \tau\bar\tau &= \boxtone + \boxttbar, \\
\boxtone \otimes \tau\bar\tau &= \boxonetbar + \boxttbar.
\end{aligned}
\ee
Since the pseudo-species anyons are a result of the fusion of the real $\tau\bar\tau$'s, they can only take value in $\{1, 1\bar\tau, \tau 1, \tau\bar\tau\}$. Fig. \ref{fig:pseudoID} depicts a solution to Eq. \eqref{eq:fusionEqDisk}. In Fig. \ref{fig:pseudoID}, exchanging $\tau 1$ and $1\bar\tau$ is the other solution, which is equivalent as it is just a matter of choice. %

\section{Pseudo-species}\label{sec:pseudo}
After the arduous journey in the previous sections, one may wonder: why bother? Simply from the fusion rules of doubled Fibonacci anyons, one knows that the anyons that can reside on an internal edge of a basis tree are $\{1,\tau 1, 1\bar\tau, \tau\bar\tau\}$. Certainly this is true. Nevertheless, the point is not to find what anyons an internal edge may represent but to figure out the meaning and significance of the pseudo-species that account for the extended anyonic exclusion statistics with fusion taken into account. The much ado in the past sections lead us to the following conceptual points.
\begin{enumerate}
\item A pseudo-species is not a single type of anyons. Rather, the single-particle states of a pseudo-species correspond to different types of anyons (see Fig. \ref{fig:pseudoID}).
This also answers the question 2) raised in the second paragraph in the introduction Section \ref{sec:intro}. Seen in Fig. \ref{fig:basisDiskWithBox}(e), a subsystem of any number of real anyons is characterized by the pseudo-species; the observables associated with the pseudo-species supply good quantum numbers (topological charges) of the subsystem. The single-particle states of the pseudo-species are precisely these topological charges. 

\item A pseudo-species may have exotic, generalized Pauli exclusion principles (see Exclusion Principle \ref{pr:excluDisk}), as compared with bosons and fermions.
\item A pseudo-species may be thought as an exotic `harmonic oscillator', which has more than one vacancy box, e.g., the pseudo-species in the previous section that has two vacancy boxes, as opposed to the harmonic oscillators of bosons and fermions. 
\item The basis in Fig. \ref{fig:boxBasisDisk} suggests a more fundamental role of pseudo-species. There, although the basis spans a multi-$\tau\bar\tau$ Hilbert space, it is solely constructed using the pseudo-species. One may be tempted to view the basis as a one-dimensional chain of pseudo-species
`oscillators'. Since exclusion statistics makes sense even in one dimension, if we take the continuum limit of the chain, we seem to obtain a field of the pseudo-species. The four types of anyons in the doubled Fibonacci order are simply different excitation modes of this field. Recall that the right most column in the basis has only one vacancy box and hence only two single-particle states, as due to the boundary effect. This fact complies with that a boundary condition affects the number of excitation modes of a quantum field, as the origin of the Casimir effect. 
\item If anyons may be the excitation modes of a fundamental field, the field cannot be real, complex, or even Grassmann number valued. In fact, in Ref.\cite{Mitra1993}, the author constructed a fundamental field theory of semions only, as certain generalized Grassman number valued field theory. Unfortunately, more general constructions, in particular of non-Abelian anyons, is yet available and is our undergoing work.
\end{enumerate}  

If we take serious the perspective that pseudo-species might be more fundamental than the physical excitations we put into the bulk, we are motivated to think of the physical anyons---or actually the fusion vertices---they bring to the basis tree (see Fig. \ref{fig:basisDiskWithBox})---as a special type of boundaries. In other words, we may treat the gapped boundaries and the physical anyons on an equal footing, namely, we may add gapped boundaries as adding physical anyons to the system. Indeed, as we shall see in the following sections.

Another important remark on the pseudo-species also regards the gapped boundary of the disk. We emphasized that it takes an entire column of two boxes to specify the single-particle states of the pseudo-species. A crucial reason behind this fact is the existence of the gapped boundary. The root of the basis tree bears only one box (or two boxes that are duplicate of each other) but there is just one pseudo-species according to $\atildedisk$; hence, if one wanted to separate the basis in Fig. \ref{fig:boxBasisDisk} into the direct product of two independent bases, one being the upper row while the other being the lower row, one would run into the trouble of decomposing the right most column of a single box. On a sphere, however, there are two pseudo-species, each taking up only one vacancy box, rendering the basis on the sphere being the direct product of two bases\cite{Hu2013}. In fact, because two disks glued along their gapped boundaries makes a sphere, we can turn our basis on a disk into the basis on a sphere in what follows.  

\begin{figure}[!h]
\centering
\includegraphics{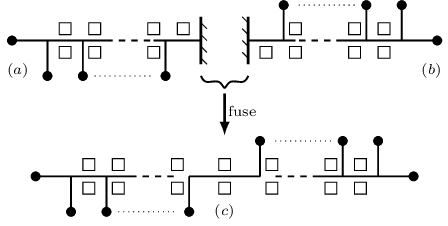}
\caption{Two bases on two disks, (a) and (b), can fuse along the boundaries of the disks into a basis on a sphere, (c).}
\label{fig:diskToSphere}
\end{figure}

We first take a generic tree basis with boxes on a disk, as in Fig. \ref{fig:diskToSphere}(a). We consider another copy of the same disk with the same basis. To glue the two disks along their boundaries, such that we can join the two bases, we need to flip the latter disk together with the basis on the disk and obtain the basis in Fig. \ref{fig:diskToSphere}(b). Then, we fuse the two bases along the boundaries. By this we mean literally treat each boundary as a composite anyon $1\oplus\tau\bar\tau$ and allow the two boundaries to fuse. This fusion produces $1, 1\bar\tau, \tau 1$, and $\tau\bar\tau$, effectively turning the two roots into an internal trunk edge, i.e., the boundaries disappear, resulting in the basis in Fig. \ref{fig:diskToSphere}(c).  

\begin{figure}[!h]
\centering
\includegraphics{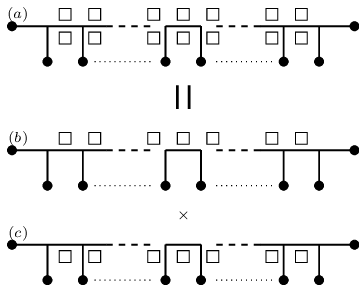}
\caption{Decompose the basis on a sphere, (a), into the direct product of two independent bases, (b) and (c). A $\bullet$ represents a $\tau\bar\tau$ as before.}
\label{fig:diskToSphereDecomp}
\end{figure}

To see how to decompose the basis in Fig. \ref{fig:diskToSphere}(c), we first deform it into that in Fig. \ref{fig:diskToSphereDecomp}(a) by bending down the upper leave in Fig. \ref{fig:diskToSphere}(c) and applying the associativity
of fusion, one at a time. 

Now we can naturally decompose the basis in Fig. \ref{fig:diskToSphereDecomp}(a) into the direct product of the two bases in Fig. \ref{fig:diskToSphereDecomp}(b) and (c). Such a decomposed basis on a sphere is indeed the one constructed in Ref.\cite{Hu2013}. Note that in the decomposed bases Fig. \ref{fig:diskToSphereDecomp}(b) and (c), a $\bullet$ still represents a $\tau\bar\tau$; yet, this does not affect the decomposition because the leaves are fixed degrees of freedom, while the true degrees of freedom reside on the trunk edges. This decomposition is mathematically well defined.
 
Moreover, after this decomposition, on a sphere, we would have two pseudo-species, respectively for the basis in Fig. \ref{fig:diskToSphereDecomp}(b) and that in Fig. \ref{fig:diskToSphereDecomp}(c). We expect to have a $3\x 3$ matrix $\tilde\alpha^{\text{sphere}}$  of mutual statistics parameters. let us index the two pseudo-species by $1$ and $2$. Since even after the basis decomposition, the Exclusion Principle \ref{pr:excluDisk}(a) and (b) still applies respectively to the row of boxes in Fig. \ref{fig:diskToSphereDecomp}(b) and the row of boxes in Fig. \ref{fig:diskToSphereDecomp}(c), we should still have $(\tilde\alpha^{\text{sphere}})_{11} =(\tilde\alpha^{\text{sphere}})_{22}=2$. Nevertheless, adding one more real anyon $\tau$ or $\bar\tau$ into the basis tree, only one box is induced on a trunk edge; hence, $(\tilde\alpha^{\text{sphere}})_{10}=(\tilde\alpha^{\text{sphere}})_{20}=-1$, i.e., half of $(\atildedisk)_{10}=-2$. We then recover the matrix \eqref{eq:alphatildeS2} on the sphere.
\section{anyonic exclusion statistics on a multi-boundary surface}\label{sec:multi}  
Having explored the extended anyonic exclusion statistics and digested all the relevant concepts and physical quantities, we are ready to gear up and study the exclusion statistics on a surface with multiple boundaries. 

\subsection{Basis construction on a multi-boundary surface}
One may suspect that we would have to run through a program similar to that in the disk case for each individual case with certain number of boundaries. In fact, however, we need not to do so. We can treat the basis we constructed on the disk as building blocks to build the new basis on a surface with a generic number of boundaries. Even stronger, via this construction of multi-boundary basis, we can directly derive or extract the statistical weight of multi-$\tau\bar\tau$ states, without doing the state counting explicitly.

To this end, we recognize that a multi-boundary surface topologically can be obtained by the connected sum of disks. A connected sum of two disks is understood as in Fig. \ref{fig:diskToCyl}. Namely, a small hole (or  a puncture, see Appendix \ref{sec:puncSphereDisk}) is removed from each of the two disks, and then the two disks are glued together along the two holes, resulting topologically a cylinder. In the topological field theory language of anyons, an anyon excitation in the bulk of a disk is attached to (or simply regarded as) a hole on the disk. So, gluing two disks, each having a $\tau\bar\tau$ in its bulk, results in a cylinder without any $\tau\bar\tau$ in its bulk, i.e., a ground state, because the two $\tau\bar\tau$'s are fused and annihilated in this procedure. Fig. \ref{fig:diskToCyl} shows a more general consideration, where in one of the disks, there are $n+1$ $\tau\bar\tau$'s, while the other has exactly one $\tau\bar\tau$, and the resultant cylinder has $n$ $\tau\bar\tau$'s.
Because the gapped boundaries are due to $\tau\bar\tau$ condensation, annihilating the two $\tau\bar\tau$'s lead to a two-state space, $\{1,\tau\bar\tau\}$, represented by an internal edge connected to the left boundary of the cylinder. This again complies with that the gapped boundary is a composite anyon $1\oplus \tau\bar\tau$.
\begin{figure}[!h]
\centering
\includegraphics[scale=0.8]{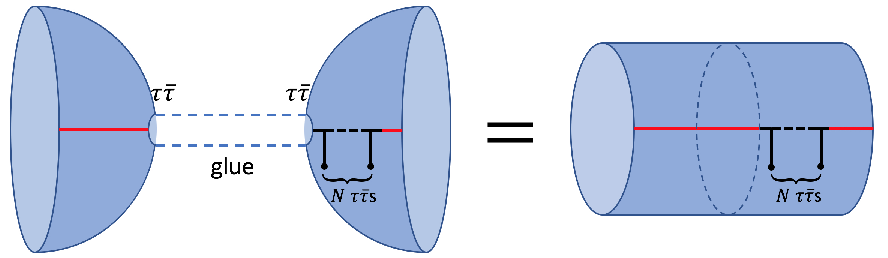}
\caption{Connected sum of two disks creates a cylinder. The disk on the left has a single hole, attached to which is a $\tau\bar\tau$. The disk on the right has $N+1$ holes, respectively attached with $N+1$ $\tau\bar\tau$'s. The resultant cylinder has $N$ $\tau\bar\tau$'s. If $N=0$, the cylinder stay in its ground state space.}
\label{fig:diskToCyl}
\end{figure}

A key point in this topological operation is that the disk with exactly one hole or $\tau\bar\tau$ can be regarded as a special unit (call it a \textit{unit disk}) that can be freely added to or removed from a system. Doing so, we can add a gapped boundary as a composite anyon to the system, on an equal footing as a real elementary anyon. As such, we can build surfaces with more than two boundaries by gluing this special unit disk to existing surfaces. In other words, the special unit disk is a new pseudo-species, as it only result in an internal edge via gluing. 

Now we have two types of real anyons, the original real anyons $\tau\bar\tau$, still indexed by $0$, and the boundary composite anyon, indexed by $1$. Warning: we used $1$ in the disk case to index the only pseudo-species there. But hereafter, $1$ indexes the boundary composite anyon. Remember that doubled Fibonacci order has only one gapped boundary type; hence, we have only one extra type of real anyons. The pseudo-species in the disk case will still be present in the case with multiple boundaries, let us index it now by $0'$.

To construct the basis with vacancy boxes on a multi-boundary surface and derive the extended statistical weight of multiple-$\tau\bar\tau$ states, let us perform the connected sum of disks in terms of the basis on a disk.
\begin{figure}[!h]
\centering
\includegraphics[scale=1.25]{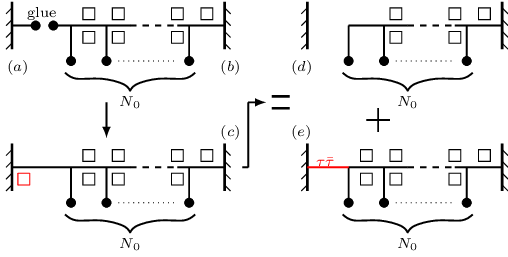}
\caption{(Color online.) The basis of $N_0$ $\tau\bar\tau$'s on a cylinder (c) obtained from gluing the basis of a single $\tau\bar\tau$ on a disk (a) to a basis of $N_0+1$ $\tau\bar\tau$'s on a disk (b) by annihilating a pair of $\tau\bar\tau$'s. The basis (c) is equivalent to the sum of the bases (d) and (e), which are in fact a basis of $N_0$ $\tau\bar\tau$'s and that of $N_0+1$ $\tau\bar\tau$'s on a disk.}
\label{fig:diskToCylBasis}
\end{figure}

Figure \ref{fig:diskToCylBasis} depicts how we obtain the basis in Fig. \ref{fig:diskToCylBasis}(c) of $N_0$ real $\tau\bar\tau$'s on a cylinder by gluing the basis in Fig. \ref{fig:diskToCylBasis}(b) of $N_0+1$ real $\tau\bar\tau$'s on a disk with the basis in Fig. \ref{fig:diskToCylBasis}(a) of the unit disk. Since we regard a gapped boundary as a composite anyon, adding such a real anyon to a disk should generate one extra internal edge---the left most edge (also a root) in Fig. \ref{fig:diskToCylBasis}(c). This new gapped boundary is also due to $\tau\bar\tau$ condensation; hence, the state on the left root can be either $1$ or $\tau\bar\tau$, indicating that one vacancy box (the red box on the left root) suffices. We index the new pseudo-species (red box) by pseudo-species $1'$, emphasizing its relation to the new gapped boundaries, in accordance with Eq. \eqref{eq:ReducedStatWeightWithBdry}. 

One can then see that the basis in Fig. \ref{fig:diskToCylBasis}(c) is equivalent to the sum of the two bases in Figs. \ref{fig:diskToCylBasis}(d) and (e). The configurations of basis in  Fig. \ref{fig:diskToCylBasis}(c) can be divided into two cases, with the left most edge either being $1$ or $\tau\bar\tau$. These two cases are equivalent to the bases in Figs. \ref{fig:diskToCylBasis}(d) and (e). In Fig. \ref{fig:diskToCylBasis}(d), the left boundary is not linked to the basis tree and thus can be forgotten, rendering the basis effectively a basis of $N_0$ $\tau\bar\tau$'s on a disk.  In Fig. \ref{fig:diskToCylBasis}(e), since the left root is deemed a $\tau\bar\tau$, the basis is effectively a basis of $N_0+1$ $\tau\bar\tau$'s on a disk. This observation has two consequences.

First, the number of states on a cylinder with $N_0$ $\tau\bar\tau$'s is the sum of the number of states on a disk with $N_0$ $\tau\bar\tau$'s and the the number of states on a disk with $N_0+1$ $\tau\bar\tau$'s on a disk. Then according to Table \ref{tab:diskFibW}, we can obtain the following table of the state counting on a cylinder.
\begin{table}[!h]
\caption{State counting of $\tau\bar\tau$ on a cylinder.}\label{tab:cylFibW}
\centering
\begin{tabular}{|m{4em}|m{2em}m{2em}m{2em}m{2em}m{2em}m{2em}m{2em}|}
\hline 
$\Ntbt$ & $0$ & $1$ & $2$ & $3$ & $4$ & $5$ & $\cdots$ \\  \hline 
$w_{P,N,C}$ & $2$ & $3$ & $7$ & $18$ & $47$ & $123$ & $\cdots$ \\ 
\hline
\end{tabular}
\end{table}
This table is also verified by computing with the extended Levin-Wen model.

Second, on a cylinder, the pseudo-species $2$ has a single-particle vacancy number $G=1$, as the new gapped boundary can be effectively not present (Fig. \ref{fig:diskToCylBasis}(d)) or effectively present (Fig. \ref{fig:diskToCylBasis}(e)). If we denote the number of new gapped boundaries by $N_1$, we would have the following new binomial factor for the corresponding pseudo-species $1'$ in the extended statistical weight to be obtained. \be\label{eq:bdryBinCyl}
\sum_{N_{1'}=0}^{1} \bpm 1\\ N_{1'} \epm,
\ee
where $N_{1'}=0$ corresponds to Fig. \ref{fig:diskToCylBasis}(d), and $N_{1'}=1$ corresponds to Fig. \ref{fig:diskToCylBasis}(e). This binomial may look rather trivial at the moment; however, we shall generalize it to cases with more than two boundaries.

Comparing Fig. \ref{fig:diskToCylBasis}(c) with (d), we see that adding one pseudo-species $1'$, i.e., the new boundary, increases the vacancy boxes of the pseudo-species $0'$ by two, while the number $N_0$ of real $\tau\bar\tau$ remain fixed. We will see shortly that this is true for any one of the new boundaries to be added. We regard pseudo-species $1'$ instead the boundary composite anyons being responsible for adding more vacancy boxes of pseudo-species $0'$ because pseudo-species $1'$ are the one that fuse with pseudo-species $0'$, seen in Fig. \ref{fig:diskToCylBasis}(c). Therefore, we have in the matrix $\atildemulti$ to be obtained, $(\atildemulti)_{0'1'}=-2$ and $(\atildemulti)_{0'1}=0$.
  
Another observation is that the number of real $\tau\bar\tau$'s does not affect the vacancy boxes of the pseudo-species $1'$. This implies that we would have $(\atildemulti)_{1'0}=0$.

\begin{figure}[!h]
\centering
\includegraphics[scale=1.25]{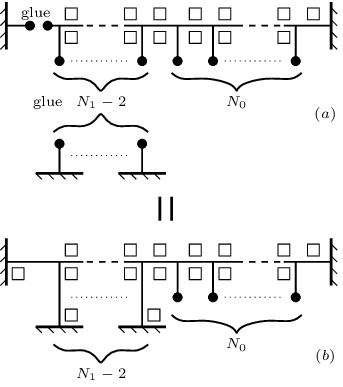}
\caption{(b) The basis of $N_0$ $\tau\bar\tau$'s on a surface with $N_1$ gapped boundaries obtained from the gluing procedure in (a). In (a), the basis of $N_0+N_1-1$ $\tau\bar\tau$'s on a disk is glued with $N_1-1$ copies of the basis of one $\tau\bar\tau$ on a disk, and $N_1-1$ pairs of $\tau\bar\tau$'s are annihilated. In (b), each new root acquires a vacancy box for the pseudo-species $2$ by the annihilation of a pair of $\tau\bar\tau$'s.}
\label{fig:diskToMultiBasis}
\end{figure}

We are now ready to construct the basis on a surface with $N_1$ gapped boundaries from the bases on a disk, as shown in Fig. \ref{fig:diskToMultiBasis}.

In Fig. \ref{fig:diskToMultiBasis}, we begin with the basis in (a) of $N_0+N_1-1$ $\tau\bar\tau$'s on a disk and glue it with $N_1-1$ copies of the basis of a single $\tau\bar\tau$ on a disk. The gluing annihilates $N_1-1$ pairs of $\tau\bar\tau$'s and results in the basis in (b) of $N_0$ $\tau\bar\tau$'s on a surface with $N_1$ gapped boundaries. In other words, we are adding $N_1-1$ composite anyons, i.e., gapped boundaries, to a disk with $N_0$ $\tau\bar\tau$'s. Note that our convention is to glue the new gapped boundaries one by one to from left to right, such that the remaining $N_0$ $\tau\bar\tau$'s are all to the right of all the new gapped boundaries. This convention is justified by the associativity of fusion interaction. 

\subsection{Statistical weight derived from the basis}

As in Fig. \ref{fig:diskToCylBasis}, annihilating each pair of $\tau\bar\tau$'s induces a vacancy box for pseudo-species $1'$ on each of the $N_1-1$ new roots in Fig. \ref{fig:diskToMultiBasis}(b). Hence, by the same argument that leads to Eq. \eqref{eq:bdryBinCyl}, if we again denote the number of pseudo-species $1'$ by $N_{1'}$, the pseudo-species $1'$ would contribute the following binomial factor to the statistical weight on the surface with $N_1$ gapped boundaries.
\be\label{eq:bdryBinMulti}
\sum_{N_{1'}=0}^{N_1-1}\bpm N_1-1 \\ N_{1'} \epm,
\ee  
which generalizes Eq. \eqref{eq:bdryBinCyl}.

Our goal is to derive the statistical weight of multi-$\tau\bar\tau$ on a surface with $N_1$ gapped boundaries, in particular, the matrix $\atildemulti$ of mutual exclusion statistics. Let us derive the entries of the matrix $\atildemulti$ first.

First of all, the number of real $\tau\bar\tau$'s is given and thus cannot be affected by any pseudo species and the number of boundaries. Besides, since each real $\tau\bar\tau$ is either present or not, we have $(\atildemulti)_{00}=1$. and $ $. Since the boundaries and the $\tau\bar\tau$'s are on an equal footing and thus can be added or removed independently, implying $(\atildemulti)_{10} =(\atildemulti)_{01}=0$.  The pseudo-species certainly cannot affect the number of real anyons and boundaries, we thus have $(\atildemulti)_{10'} = (\atildemulti)_{11'}= (\atildemulti)_{00'} = (\atildemulti)_{01'}=0$.

A tricky entry is $(\atildemulti)_{11}$. To find this one, we can consider an extreme scenario where there is no any real anyons $\tau\bar\tau$ in the bulk, such that the system is in the ground states. According to Ref. \cite{HungWan2014}, the ground-state Hilbert space is nontrivial and has a degeneracy. This degeneracy is actually a result of the fusion of the pseudo-species $1'$ but not of the boundary composite anyons. In other words, the number of gapped boundaries does not affect the state counting directly but instead via its associated pseudo-species $1'$. Thus, we can conclude that $(\atildemulti)_{11}=0$. By the same token, $G_1=1$ must hold. 

In the previous subsection, we already know that $(\atildemulti)_{0'1'}=-2$ and $(\atildemulti)_{0'1}=0$. Also from Fig. \ref{fig:diskToCylBasis}, we can see that the way how a real $\tau\bar\tau$ modifies the number of vacancy boxes of pseudo-species $0'$ on a disk still holds on a multi-boundary surface. Namely, one extra $\tau\bar\tau$ again adds two vacancy boxes for pseudo-species $0'$ in the same way as it does on a disk. Hence, $(\atildemulti)_{0'0}=-2$. By the same token, the exclusion principle on the vacancy boxes for pseudo-species $0'$ on a disk remains true on a multi-boundary surface, meaning $(\atildemulti)_{0'0'}=2$.

As reasoned below Figs. \ref{fig:diskToCylBasis} and \ref{fig:diskToMultiBasis}, each pseudo-species $1'$ takes up only one vacancy box that is either occupied or unoccupied. We thus have $(\atildemulti)_{1'1'}=1$. According to Fig. \ref{fig:diskToMultiBasis}(b), each gapped boundary, which is indexed by $1$, grants one vacancy box for pseudo-species $1'$, hence we have $(\atildemulti)_{1'1}=-1$; The real $\tau\bar\tau$'s do not affect the vacancy boxes of pseudo-species $1'$; hence, $(\atildemulti)_{1'0}=0$. The only missing entry now is $(\atildemulti)_{1'0'}$, which can be found by the binomial factor \eqref{eq:bdryBinMulti}. The $N_1-1$ in the binomial should be understood as $\tilde G_{0'}^{eff}=N_1-1$. Following the definition in Eq. \ref{eq:StatWeightMulti}, we have
\begin{equation*}
\begin{aligned}
N_1-1 = \tilde G_{1'}^{eff} &= \tilde G_{1'} -\sum_j (\atildemulti)_{1'j} (N_j - \delta_{1'j}) + (N_{1'} - 1) \\
&= \tilde G_{1'} - (\atildemulti)_{1'1} N_1, 
\end{aligned}
\end{equation*}
where use of $(\atildemulti)_{1'0}=0$, $(\atildemulti)_{1'1}=-1$, and $(\atildemulti)_{1'1'}=1$ is made to get the third equality. Again from Fig. \ref{fig:diskToMultiBasis}(b), we can see that whether a pseudo-species $1'$, i.e., a new gapped boundary, is present or not has nothing to do with the real species or pseudo-species $0'$. In other words, $\tilde G_{1'}=-1$ as well, which is the very $-1$ in the binomial of Eq. \eqref{eq:bdryBinMulti}. Plugging this into the equation above, we conclude that $(\atildemulti)_{1'0'}=0$.

We therefore obtain the full $\tilde\alpha$ matrix on a surface with $M_1$ boundaries:
\be\label{eq:alphaMultiBdry}
\atildemulti=\bordermatrix{
& 0 &1 &0'&1'\cr
0& 1 & 0 & 0&0\cr
1& 0&0&0&0\cr
0'& -2 &0& 2 & -2\cr
1'&0 &-1& 0 & 1
}.
\ee

The final task is to find $\tilde G_{0'}$ in the case with $N_1$ gapped boundaries. Keeping the matrix \eqref{eq:alphaMultiBdry} in mind, we have
\begin{equation*}
\begin{aligned}
\tilde G_{0'}^{eff}  =& \tilde G_{0'} -\sum_{j=0}^2 (\atildemulti)_{0'j} (N_j - \delta_{{0'}j}) + (N_{0'} - 1) \\
 =& \tilde G_{0'} + 2 N_0 + 2N_{1'} - (N_{0'}-1)\\
 \xeq{N_{0'}=1} & \tilde G_{0'}  + 2 N_0 + 2N_{1'},
\end{aligned}
\end{equation*}
where the third equality holds for $N_{0'}=1$ because we are interested in the single-particle state number of pseudo-species $1$ only. The equation above should reduce to the disk case when $N_1=1$, i.e., $N_{1'}=0$, where we have $\tilde G_{0'}^{eff} = \tilde G_{0'} + 2 N_0$. Nevertheless, we have already learnt that on a disk, $\tilde G_{0'} = -3$, which now should hold for the $N_1$-boundary surface too. 

Above all, we can extract the full statistical weight of the multi-$\tau\bar\tau$ space on a surface with $N_1$ boundaries. Namely,
\be\label{eq:statWeightMultiBdry}
\begin{aligned}
W_{P,N_{0}}^{N_1} &=  \bpm P \\ N_0 \epm \bpm N_1 \\ N_1 \epm
\sum_{N_{1'}=0}^{N_1-1} \bpm N_1-1 \\ N_{1'} \epm  \sum_{N_{0'}=0}^{N_0+N_{1'}-1} \bpm
   2 N_0 + 2 N_{1'} - N_{0'} - 2 \\ N_{0'} \epm\\
& =  \bpm P \\ N_0 \epm 
\sum_{N_{1'}=0}^{N_1-1} \bpm N_1-1 \\ N_{1'} \epm  \sum_{N_{0'}=0}^{N_0+N_{1'}-1} \bpm
   2 N_0 + 2 N_{1'} - N_{0'} - 2 \\ N_{0'} \epm .  
\end{aligned}   
\ee
The first binomial factor above is our assumption that there are $P$ vacancy boxes (such as the plaquettes in the extended Levin-Wen model) for the real $\tau\bar\tau$'s. The upper limit of the second summation above is the total number of internal edges for pseudo-species $0'$ in the basis, which can be read off from Fig. \ref{fig:diskToMultiBasis}(b). Equation \eqref{eq:statWeightMultiBdry} is clearly an instance of our proposed general formula \eqref{eq:ReducedStatWeightWithBdry}.

In the second line of Eq. \eqref{eq:statWeightMultiBdry}, we dropped the trivial binomial factor $(\begin{smallmatrix}N_1 \\ N_1 \end{smallmatrix}) \equiv 1$. This does not mean we can naively drop the second row in the matrix \eqref{eq:alphaMultiBdry} because dropping this row would drop the second column too but $(\atildemulti)_{1'1}=-1$, i.e., the second column cannot be dropped. Clearly, when $N_1=1$, which forces $N_{1'}=0$, the matrix \eqref{eq:alphaMultiBdry} and the statistical weight \eqref{eq:statWeightMultiBdry} reduce to the matrix \eqref{eq:alphaDisk} and Eq. \eqref{eq:WdiskBin} in the disk case. This further verifies our general results.

\begin{figure}[!h]
\centering
\includegraphics[scale=0.8]{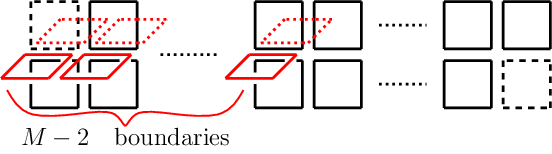}
\caption{(Color online.) Simplified basis of a multi-$\tau\bar\tau$ Hilbert space on a surface with $N_1$ gapped boundaries. This is three dimensional picture. The vertical columns with solid boxes and the right most column are vacancy boxes for pseudo-species $0'$, as in the disk case. The left most vertical column with one solid and one dashed box and the horizontal columns (in red) are the vacancy boxes for pseudo-species $1'$. Duplicate boxes (dashed) are manifest.}
\label{fig:boxBasisMultiBdry}
\end{figure}

With the statistical weight \eqref{eq:statWeightMultiBdry} and the matrix \eqref{eq:alphaMultiBdry} at our disposal, we can directly write down the generalized Pauli exclusion principle on a surface with $N_1$ gapped boundaries. Before we do this, let us first draw the simplified basis consisting only vacancy boxes in Fig. \ref{fig:boxBasisMultiBdry}. The basis is naturally presented in a $3$-dimensional form because of the existence of pseudo-species $1'$. Compared to Fig. \ref{fig:diskToMultiBasis}(b), our convention is to put the box on the left most root in Fig. \ref{fig:diskToMultiBasis}(b) as the left most vertical column in Fig. \ref{fig:boxBasisMultiBdry}, then draw the box on each of the $N_1-2$ roots in \ref{fig:diskToMultiBasis}(b) as a horizontal column (in red) perpendicular to the vertical column of pseudo-species $0'$ boxes to the immediate right of the root.    

Since pseudo-species $1'$ is a new species different from pseudo-species $0'$, the exclusion principle on this species anyons is better presented when the vacancy boxes for them are next to one another. As such, there is no other way of drawing the boxed basis more natural than the $3$-dimensional one in Fig. \ref{fig:boxBasisMultiBdry}. 

The boxed basis in Fig. \ref{fig:boxBasisMultiBdry} entitles a generalized exclusion principle presented in the following simple form.
\begin{principle}
\begin{equation*}
\includegraphics[scale=0.8]{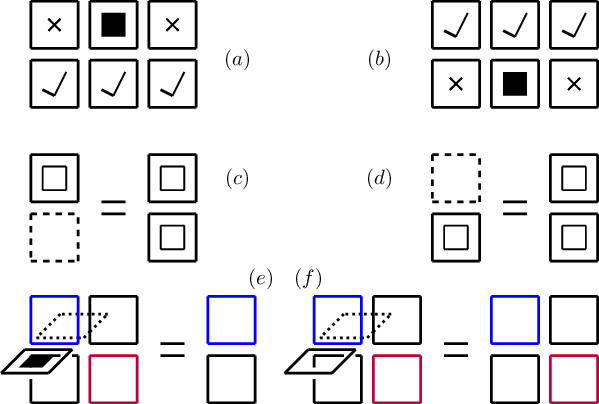}
\end{equation*}
Here, rules (a) through (d) are in fact the same rules on a disk. Rules (e) and (f) are new; their function is to reduce a $3$-dimensional situation to $2$-dimensional, such that the rules (a) through (d) can be directly applied. A blue box can be either a solid or dashed box. So is a purple box. The other conventions for understanding the rules are the same as those in Exclusion Principle \ref{pr:excluDisk}.
\end{principle}   

The identification of pseudo-species $1$ anyons in Fig. \ref{fig:pseudoID} remains true in the multi-boundary case. Pseudo-species $2$ anyons are simply the first two in Fig. \ref{fig:pseudoID}.

\section{Equivalence relation between statistical weight and fusion basis}
\label{sec:equiv}

In Sections \ref{sec:Disk} and \ref{sec:basisDisk}, we have shown how we may construct the basis of multi-$\tau\bar\tau$ Hilbert space on a disk and obtain the generalized exclusion principle on a disk by understanding the matrix of mutual exclusion statistics on a disk. In Section \ref{sec:multi}, we have marched in the opposite direction. Namely, we directly construct the basis of multi-$\tau\bar\tau$ space on a multi-boundary surface by using the bases on a disk as building blocks. Along with this construction, we derive explicitly the statistical weight and prove that the generalized Pauli exclusion principle holds. We derive explicitly the exclusion statistics parameters on the multi-boundary surface. Therefore, both the extended anyonic exclusion statistics and the basis of a multi-$\tau\bar\tau$ space characterize the Hilbert space structure completely and equivalently. Since the sphere case can be obtained from the disk case by the topological operation introduced in Section \ref{sec:pseudo}, we have a more general equivalence relation as follows.
\be\label{eq:equiv}
\text{anyonic exclusion statistics} \Longleftrightarrow \text{Fusion Basis.}
\ee

\begin{figure}[!h]
        \centering
        \includegraphics[scale=0.5]{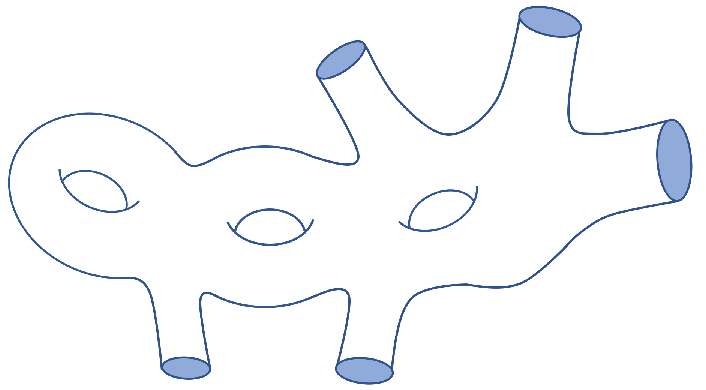}
        \caption{A genus-$3$ Riemann surface with five boundaries.}
        \label{fig:genusBdry}
\end{figure}

So far, we have only considered genus-$0$ surfaces with gapped boundaries. In fact, by the topological operations in Fig. \ref{fig:diskToMultiBasis} and that in Fig. \ref{fig:diskToSphere}, we can obtain the basis of a multi-$\tau\bar\tau$ space on any (even self-knotted) Riemann surface with nonzero genus number and multiple gapped boundaries, e.g., Fig. \ref{fig:genusBdry}, and hence the corresponding extended exclusion statistics. 

The topological operations also work in the cases with Chiral topological orders with gapless boundaries. In such cases, one can start with the basis of real anyons on a sphere (recall Fig. \ref{fig:diskToSphereDecomp} and the punctured sphere picture in Appendix \ref{sec:puncSphereDisk}) and use the topological gluing to obtain the bases on any Riemann surfaces.

We conjecture that the equivalence between the anyonic exclusion statistics and the basis of the corresponding Hilbert space not only holds for the doubled Fibonacci anyons but also holds in a system of anyons described by a modular tensor category. In a companion paper in preparation, we try to prove this conjecture. In Appendix \ref{sec:multiRealSpecies}, we considered as an example the case where both $\tau\bar\tau$ and $1\bar\tau$ are real anyons on a disk, no violation of the equivalence relation. In Appendix \ref{sec:Z2TC}, we also studied the case of Abelian anyons, using the $\Z_2$ Toric Code topological order as an example, and no violation of the equivalence either.   

The equivalence relation \eqref{eq:equiv} enables one to obtain a canonical set of exclusion statistics parameters, up to possible direct product decomposition, such as that in Fig. \ref{fig:diskToSphereDecomp}, and associativity of fusion. If one tries to find a binomial fitting of a state counting, the fitting is by no means unique, not even unique up to decomposition. In obtaining the statistical weight \eqref{eq:WdiskBin} and $\atildedisk$ \eqref{eq:alphaDisk} from the state counting in Table \ref{tab:diskFibW}, we actually just chose the smallest possible matrix out of the various possible
binomial fittings of the table. The choice just fortunately turns out to be the canonical one that can be directly derived from the corresponding basis construction.

Note that if one applied associativity of fusion to a basis, one would obtain a different basis. The old basis and the new basis are related by a linear transformation. However, the two bases result in the same $\tilde\alpha$ matrices.

\acknowledgments
We appreciate Yong-shi Wu for illuminating discussions, comments, and suggestions. We also thank Ling-Yan Hung for helpful discussions. YDW thanks IQC for hospitality during his visit, where this paper is finalized; he is also supported by the Shanghai Pujiang Program No. KBH1512328. 

\appendix
\section{The Extended Levin-Wen Model}\label{appd:LW}
In this appendix, we briefly review the key ingredients of the extended Levin-Wen model that are necessary for understanding our method in Section \ref{sec:Disk}.

The model is defined by an exactly solvable Hamiltonian defined on a trivalent lattice embedded in a surface with boundaries, e.g., the lattice in Fig. \ref{fig:LWdisk}. The Hamiltonian of the model reads
\be\label{eq:appdLWHam}
H=H_{\mathrm{bulk}}+H_{\mathrm{bdry}},
\ee
with
\be\label{eq:appdBulkHam}
H_\mathrm{bulk}=-\sum_{v\in \mathrm{bulk}} A_v -\sum_{p\in \mathrm{bulk}} B_p,
\ee 
and
\be\label{eq:appdBdryHam}
H_\mathrm{bdry}=-\sum_{v\in \mathrm{bdry}} \bar A_v -\sum_{p\in \mathrm{bdry}} \bar B_p,
\ee
where the sums run over the vertices and plaquettes of the lattice respectively.

The degrees of freedom of the model live  on the edges of the lattice. The bulk degrees of freedom take value in (simple representatives of) objects of a UFC, while the boundary degrees of freedom take value in a Frobenius algebra object of the UFC. This is explained in Section \ref{subsec:eLWfluxon}. The bulk operators in the Hamiltonian \eqref{eq:appdBulkHam} are defined as follows. The operator $A_v$ acts on a vertex $v$ and imposes the fusion rule on the vertex:
\begin{align}
A_v\BLvert\vertex\Brangle=\delta_{ijk}\BLvert\vertex\Brangle,
\end{align}
where
\be
\delta_{ijk}=1\quad\mathrm{if}\quad i \otimes j \otimes k\ni 0. 
\ee
The 
Operator $B_p$ acts on the plaquette $p$ and is a composite operator:
\begin{align}
B_p=\frac{1}{D}\sum_{s\in L}d_s B_p^s,
\end{align}
where
\begin{align}
&B_p^s\BLvert\plqt\Brangle=\\
&v_{j_1}v_{j_2}v_{j_3}v_{j_1'}v_{j_2'}v_{j_3'}G^{j_4j_3^*j_1}_{s^*j_1'{j'_3}^*}G^{j_5j_1^*j_2}_{s^*j_2'{j'_1}^*}G^{j_6j_2^*j_3}_{s^*j_3'{j'_2}^*}\BLvert\plqtp\Brangle.
\end{align}
Here $v_i=\sqrt{d_i}$, and $d_i$ is the quantum dimension for anyon type $i$. A factor $G$ is called a 6$j$-symbol, which is a solution to the following conditions given the set $L$ and the quantum dimensions,
\begin{align}
G^{ijm}_{kln}&=G^{mij}_{nk^*l^*}=G^{klm^*}_{ijn^*}=\mathrm{sgn}(d_m)\mathrm{sgn}(d_n)\bar G^{j^*i^*m^*}_{l^*k^*n},\notag \\
&\sum_n d_n G^{mlq}_{kp^*n}G^{jip}_{mns^*}G^{js^*n}_{lkr^*}=G^{jip}_{q^*kr^*}G^{riq^*}_{mls^*},\\
&\sum_n d_n G^{mlq}_{kp^*n}G^{l^*m^*i^*}_{pk^*n}=\frac{\delta_iq}{d_i}\delta_{mlq}\delta_{k^*ip}.\notag
\end{align}
Both of these two types of operators are projectors and commute with each other, namely,
\be
[B_p,B_{p'}]=[A_v,A_{v'}]=0=[B_p,A_v].
\ee

The boundary Hamiltonian is constructed similarly. Operator $\bar A_v$ projects the boundary degrees of freedom from a UFC to a Frobenius algebra, namely,
\begin{align}
\bar A_v \BLvert\bvertex\Brangle=\delta_{a_n\in A} \BLvert\bvertex\Brangle,
\end{align}
and operator $\bar B_p$ is again a composite operator as
\be
\bar B_p=\frac{1}{DA}\sum_{t\in A}v_t \bar B_p^t,\quad D_A=\sum_{t\in A}d_t.
\ee
The $\bar B_p^t$ operator acts on a boundary open plaquette as
\begin{align}
\bar B_p^t \BLvert\bplqt\Brangle=\sum_{a'_1,a'_2,j'_2,j'_3}\notag
f_{t^*{a'_2}^*a_2}f_{t^*{a'_1}^*a_1}u_{a_1}u_{a_2}u_{a'_1}u_{a'_2}\\
G^{j_4^*j_3a_2^*}_{t^*{a'_2}^*j'_3}G^{j_5j_2j_3^*}_{t^*{j'_3}^*j'_2}G^{j_1a_1^*j_2}_{t^*{j'_2}{a'_1}^*}v_{j_2}v_{j_3}v_{j'_2}v_{j'_3}\BLvert\bplqtp\Brangle,
\end{align}
where $u_a=\sqrt{v_a}$, and the coefficients $f_{ijm}$ are defined by the multiplication of  the algebra $A$, i.e.,
\be
\sum_c f_{abc^*}f_{cde*}G^{abc^*}_{de^*g}v_cv_g=f_{age^*}f_{bdg^*},
\ee
and satisfy
\begin{align}
f_{bb^*0}=1, \quad b\in A.\\
f_{abc}=f_{cab},\\
\sum_{ab}f_{abc}f_{c^*b^*a^*}v_av_b=d_Av_c.
\end{align}
The boundary operators are also projectors, commute with each other, and commute with the bulk operators. The total Hamiltonian \ref{eq:appdLWHam} is thus exactly solvable. The ground states satisfy the condition $A_v=B_p=1$ for all $v$ and $p$, while excited states violate these constraints for certain vertex or plaquette. For more details of extended LW model, one is referred to Ref \cite{Hu2017a}.  

\section{Punctured Sphere versus Disk}\label{sec:puncSphereDisk}

\begin{figure}
\centering
\subfigure[]{
\includegraphics[scale=0.45]{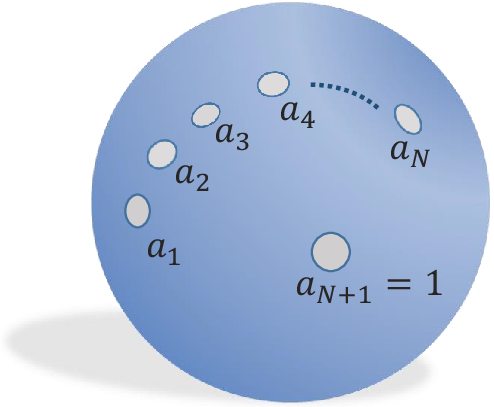}\label{fig:puncSphere}
}
\subfigure[]{
\includegraphics[scale=0.45]{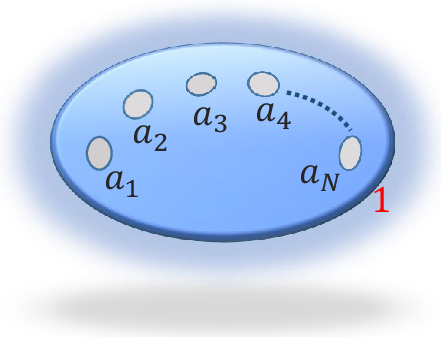}\label{fig:puncDisk}
}
\caption{(a) An (N+1)-puncture sphere. (b) A disk with $N$ punctures.}
\end{figure}

In a Fibonacci system, a logical qubit is realized as subspaces of the Hilbert space of three Fibonacci anyons, and more logical qubits require more Fibonacci anyons. The Hilbert space of a number of Fibonacci anyons is understood as the fusion space of these anyons. Nevertheless, such fusion spaces are obtained in an abstract fashion, namely the punctured-sphere picture. In this picture, a sphere with $N+1$ punctures can inhabit $N$ Fibonacci anyons respectively on $N$ of the punctures, whereas on the $(N+1)$-th puncture there resides the vacuum $1$ (see Fig. \ref{fig:puncSphere}). In other words, the $N$ Fibonacci anyons must fuse to the vacuum. Since the $N+1$ puncture has nothing, it is thought to be equivalent to an empty hole.
 
Similarly, on a disk with a gapped boundary, we can also have a punctured-disk picture of $N$ $\tau\bar\tau$'s, as in Fig. \ref{fig:puncDisk}. The $N$ $\tau\bar\tau$'s in the bulk of the disk must fuse as a lump into an anyon that can connect to the gapped boundary. This anyon, according to \cite{HungWan2014}, is either the trivial anyon $1$ or $\tau\bar\tau$ (see Fig. \ref{fig:diskBasis}). Since the boundaries are treated as composite anyons, it thus renders the disk actually a sphere with the $N+1$-th anyon a composite one.

Both the punctured-sphere picture and the punctured-disk picture allow one to glue multiple spheres or disks by identifying certain pairs of punctures to form more complex Riemann surfaces with or without boundaries. Identifying a pair of punctures physically means fusing the pair of $\tau\bar\tau$'s attached to the pair of punctures. 

\section{More Than One Real Species on a Disk.}\label{sec:multiRealSpecies}
\begin{figure}[!h]
\centering
\includegraphics[scale=1.25]{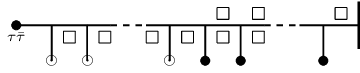}
\caption{The basis when both $\tau\bar\tau$ ($\bullet$) and $1\bar\tau$ ($\odot$) are real anyons in the bulk. Our convention here is to put a $\tau\bar\tau$ on the leftmost, followed by all $1\bar\tau$'s, and then other $\tau\bar\tau$'s adjacent to the boundary. The convention we take is natural, as a single $1\bar\tau$ cannot exist on a disk without any $\tau\bar\tau$.}
\label{fig:twoSpe}
\end{figure}

In this appendix, we allow both $\tau\bar\tau$ and $1\bar\tau$ to exist as real anyons on a disk. Table \ref{tab:2speciesDisk} records the state counting for a few numbers of $\tau\bar\tau$'s and $1\bar\tau$'s.
\begin{table}[h]
\caption{State counting on a disk with both real $\tau\bar\tau$'s and $1\bar\tau$'s.}
\centering
\begin{tabular}{|p{1cm}|p{1cm}p{1cm}p{1cm}p{1cm}|}
\hline
\diagbox{$\tau\bar\tau$}{$1\bar\tau$} & 0 & 1 & 2 & 3 \\
\hline
0 & 1 & 0 & 1 & 1\\
1 & 1 & 1 & 2 & 3\\
2 & 2 & 3 & 5 & 8\\
3 & 5 & 8 &13& 21\\
4 &13&21&34&55\\
\hline
\end{tabular}
\label{tab:2speciesDisk}
\end{table}
We can consequently obtain the statistical weight as follows by binomial fitting of the Table
\begin{align}
W_{\{N_{\tau\bar\tau},G_{\tau\bar\tau}|N_{1\bar\tau},G_{1\bar\tau}\}}=\left(\begin{array}{c} P_1 \\ N_{\tau\bar\tau}\end{array}\right)\left(\begin{array}{c} P_2 \\ N_{1\bar\tau}\end{array}\right)
\sum_{N_1}^{\frac{1}{2}(2N_{\tau\bar\tau}+N_{1\bar\tau}-2)}\left( \begin{array}{c} 2N_{\tau\bar\tau}+N_{1\bar\tau}-N_1-2 \\ N_1 \end{array}\right).
\end{align}
The corresponding matrix of mutual exclusion statistics is then
\begin{align}\label{1taubaralpha}
\tilde\alpha=\left(\begin{array}{ccc} 1 & 0 & 0\\ 0 & 1 & 0\\-2 & -1 & 2 \end{array}\right).
\end{align}
In the matrix, we index $\tau\bar\tau$ by $0$, $1\bar\tau$ by $1$, and the pseudo-species by $2$. Real $\tau\bar\tau$'s affect the pseudo-species vacancies in the same way as they do in the case where only $\tau\bar\tau$'s are allowed to be real anyons, as $\tilde\alpha_{20}=-2$. An extra real $1\bar\tau$ however only adds one vacancy box for the pseudo-species because $\tilde\alpha_{21}=-1$. The exclusion principle on the pseudo-species is the same as in the case where only $\tau\bar\tau$'s are real  because $\tilde\alpha_{22}=2$. Note that although the matrix above is $3\x 3$, since there is only one pseudo-species, the boxed bases of the corresponding Hilbert spaces is only $2$-dimensional. Fig. \ref{fig:twoSpe} depicts the basis tree.

The entry $\tilde\alpha_{21}=-1$ is new, and as seen in Fig. \ref{fig:twoSpe}, each additional $1\bar\tau$ adds one vacancy box for the pseudo-species. The vacancy boxes due to adding $1\bar\tau$'s are below the trunk of the tree basis and result in an exclusion rule (Fig. \ref{fig:twoRule}) in addition to the generalized Pauli exclusion principle \ref{pr:excluDisk}. 
\begin{figure}[!h]
\centering
\twoSpeciesRule
\caption{New rule of both $\tau\bar\tau$ and $1\bar\tau$ on a disk. The $\boxed{\square}$ is still a wildcard. The rule indicates that the upper box in the same column with the box added by $1\bar\tau$ is always unoccupied.}
\label{fig:twoRule}
\end{figure}

\section{$\Z_2$ Toric Code on a Disk}\label{sec:Z2TC}
In this appendix, we look at the $\Z_2$ toric code topological order as an example of the case of Abelian anyons. Since the fusion between any two Abelian anyons produces exactly one anyon, which is also Abelian, any internal edge in any fusion basis of a multi-anyon Hilbert space has a fixed degree of freedom. Hence, there is no pseudo-species and thus no $\tilde\alpha$ matrix in Abelian cases. Nevertheless, the exclusion statistics in Abelian cases is still interesting, in particular when gapped boundaries are present.

The $\Z_2$ toric code topological order has four anyons $\{1,e,m,\epsilon\}$ with the nontrivial fusions $e\ox e = m\ox m = \epsilon\ox \epsilon =1$, $e\ox m= \epsilon$, $e\ox \epsilon = m$, and $m\ox \epsilon = e$. On a sphere, $e$ and $m$ are have trivial self exchange statistics, i.e., they are self-bosons; however, they are mutually semions. An anyon $\epsilon$ is a self-fermion. On a sphere, the states of odd number of anyons of any species when other species anyons are not present do not exist.

On a disk, this topological order has two possible gapped boundary conditions, the $e$-boundary and $m$-boundary respectively due to condensing $e$ and condensing $m$ at the boundary. 

Let us consider a disk with the $e$ boundary, on which only $e$ can move to the boundary. Clearly, the number of states for any number of real $e$'s in the bulk of the disk is identically unity. Thus, the statistical weight for $N_e$ $e$'s on a disk is simply $(\begin{smallmatrix}P\\N_e \end{smallmatrix})$, indicating that $e$ is simply a hardcore boson, as a consequence of the $e$-boundary. 

If we count the number of states of electric charge $e$ without $m$ and $\epsilon$ via Kitaev model, what we get is a sequence of 1. That is, the electric charge $e$ becomes a hard core boson just like the trivial anyon. Furthermore, we find the state counting for both $m$ and $\epsilon$ yields the same sequence.

\begin{table}[!h]
        \caption{Counting states with any number of $m$ and $\epsilon$.}
        \label{tab:m_epsilonDiskToric}
        \centering
        \begin{tabular}{|p{1cm}|p{1cm}p{1cm}p{1cm}p{1cm}p{1cm}p{1cm}|}
                \hline
                \diagbox{$\epsilon$}{m} & 0 & 1 & 2 & 3 & 4 & 5\\
                \hline
                0 & 1 & 0 & 1 & 0 & 1 & 0\\
                1 & 0 & 1 & 0 & 1 & 0 & 1\\
                2 & 1 & 0 & 1 & 0 & 1 & 0\\
                3 & 0 & 1 & 0 & 1 & 0 & 1\\
                4 & 1 & 0 & 1 & 0 & 1 & 0\\
                5 & 0 & 1 & 0 & 1 & 0 & 1\\
                \hline
        \end{tabular}
\end{table} 

We can also count the states for coexisting $m$'s and $\epsilon$'s without the presence of $e$ in the bulk. Table \ref{tab:m_epsilonDiskToric} captures the counting. Recall that on a sphere, a pair of $m$ and $\epsilon$ cannot exist without an $e$; however, on a disk, it is allowed, so are similar states, as seen in the table. Since the (fusion) Hilbert space dimension is either $1$ or $0$, corresponding to a $1$-dimensional basis or no basis at all, the equivalence relation \eqref{eq:equiv} trivially holds, as pointed out in Section \ref{sec:equiv}.


\bibliographystyle{apsrev4-1}
\bibliography{StringNet}
\end{document}